\documentclass[]{article}
\usepackage[]{arxiv}
\usepackage{svg}
\usepackage[utf8]{inputenc} 
\usepackage[T1]{fontenc}    
\usepackage{hyperref}       
\usepackage{url}            
\usepackage{booktabs}       
\usepackage{amsfonts}       
\usepackage{amsmath}
\usepackage{nicefrac}       
\usepackage{microtype}      
\usepackage{lipsum}		
\usepackage[linesnumbered, ruled]{algorithm2e}
\usepackage{colortbl}
\usepackage{multirow}
\usepackage{graphicx}

\author{Lloyd Fung \thanks{corresponding email: lsf27@cam.ac.uk} \\
Department of Applied Mathematics and Theoretical Physics\\ University of Cambridge \\ Cambridge, CB2, UK \\
	\And
	Urban Fasel \\
	Department of Aeronautics \\ Imperial College London \\ London, SW7 2AZ, UK\\
	\AND
	Matthew Juniper\\
	Cambridge University Engineering Department \\ Cambridge, CB2 1PZ, UK\\
}

\title{Rapid Bayesian identification of sparse nonlinear dynamics from
scarce and noisy data}

\begin{document}
\maketitle
\begin{abstract}
  We propose a fast probabilistic framework for identifying differential
  equations governing the dynamics of observed data.  We recast the
  SINDy method within a Bayesian framework and use Gaussian
  approximations for the prior and likelihood to speed up computation. The
  resulting method, Bayesian-SINDy, not only quantifies uncertainty in the
  parameters estimated but also is more robust when learning the correct model
  from limited and noisy data. Using both synthetic and real-life examples such as Lynx-Hare population dynamics, we demonstrate the
  effectiveness of the new framework in learning correct model equations
  and compare its computational and data efficiency with existing methods.
  Because Bayesian-SINDy can quickly assimilate data and is robust against noise,
  it is particularly suitable for biological data and real-time system
  identification in control. Its probabilistic framework also enables the
  calculation of information entropy, laying the foundation for an active learning strategy.
\end{abstract}

 
    
\maketitle


\hypertarget{sec:intro}{%
\section{Introduction}\label{sec:intro}}

The pursuit of direct model equation discovery has been an ongoing and
significant area of interest in scientific machine learning. The popular
\emph{sparse identification of nonlinear dynamics} (SINDy) framework
\cite{Brunton2016} offers a promising approach to extract parsimonious
equations directly from data. SINDy's promotion of parsimony by sparse regression allows for the identification of an interpretable model that
balances accuracy with generalizability, while its simplicity leads to a
relatively efficient and fast learning process compared to other machine
learning techniques. 
The framework has been successfully applied in a variety
of applications, such as model idenficiation in plasma physics \cite{kaptanoglu2021physics}, control engineering
\cite{Fasel2021, Fuentes2021}, biological transport problems \cite{lagergren2020learning}, socio-cognitive systems \cite{dale2018equations}, epidemiology \cite{jiang2021modeling,Horrocks2020} and turbulence modelling
\cite{ChuaKhoo2022}.  Furthermore, its remarkable extendibility has attracted a
range of modifications, including the adaptation to discover partial
differential equations \cite{Rudy2017}, the extension to libraries of
rational functions \cite{Kaheman2020}, the integration of ensembling
techniques to improve data efficiency \cite{Fasel2022} and the use of
weak formulations \cite{Messenger2021, Reinbold2020} to avoid noise
amplification when computing derivatives from discrete data.

One major difficulty in using scientific machine learning methods in fields 
such as biophysics, ecology, and microbiology, is that measured data from these fields is often noisy and scarce.
This limitation renders many data-intensive machine learning techniques infeasible.
Learning speed is also a major requirement in other real-time applications, such as control.
Although SINDy is less data-intensive and faster than most machine learning techniques, it is
nonetheless susceptible to noise when the dataset is scarce. 
In some cases, SINDy may have
difficulties recovering accurate and parsimonious models if the data is
too scarce and noisy. To address the challenge of fast and robust learning in a
noisy and scarce dataset, we present Bayesian-SINDy. This method
combines sparse Bayesian learning and SINDy to extract parsimonious
governing equations directly from the data. It accounts for the uncertainty
in the data and the estimated parameters when selecting parsimonious models, 
but does so without the need for
expensive sampling computations such as Markov Chain Monte Carlo (MCMC),
making it particularly suitable for real-time applications in which fast
computation is required.

To appreciate the potential improvement of SINDy's performance in scarce
and noisy data using sparse Bayesian learning, one must first understand the principles of SINDy. Essentially, SINDy approaches
equation discovery as a sparse dictionary learning problem. First, it
constructs a dictionary consisting of all potential candidate functions
of the input data. Then, SINDy employs sparse regression to identify the
most pertinent functions that describe the dynamics and to eliminate the
unnecessary ones, yielding parsimonious dynamical equations that retain
only the relevant terms. The original study \cite{Brunton2016}
established the Sequentially Thresholded Least Squares (STLS) algorithm as a
method for sparse regression with minimal computational cost. STLS
assumes that all relevant terms possess coefficients exceeding a certain
threshold, allowing for the elimination of irrelevant terms with coefficients below this threshold. 
Hence, the imposed threshold is effectively a
hyperparameter one can tune to control how aggressively sparsity is
promoted. Besides STLS, there are other regularisation and
relaxation techniques that promote sparsity during regression, such as
LASSO \cite{tibshirani1996regression} and SR3 \cite{zheng2019unified}. Nevertheless, nearly all sparse regression techniques contain
hyperparameters that require fine-tuning. To properly select
hyperparameters, the Pareto front is commonly employed as a heuristic
method to ascertain the appropriate degree of sparsity. However,
interpreting the front can be ambiguous, especially in
cases where the front lacks a distinct elbow point. The later introduction of Akaike/Bayesian
information criterion (AIC/BIC) in the model selection process \cite{Mangan2017} provides
certain remedies to the problem, but they only work well in the high data
limit \cite{Burnham2004}.

Another prevalent problem with existing SINDy-like algorithms is their
ability to handle scarce and noisy data. While SINDy-based approaches
are effective at identifying equations from synthetic data with little
noise, real-world measurements are typically scarce and prone to
significant measurement error. Large noise in the data presents several
challenges to SINDy. 
First, measurement noise corrupts the precise recovery of time derivatives 
because conventional finite difference approximations can amplify noise by the
order of the sample rate. This problem can be mitigated by using a weak
formulation \cite{Messenger2021, Reinbold2020}. 
However, this approach may also introduce bias to the data, as
we will illustrate in \S\ref{sec:WeakVSFD}. Second, excessive noise in
the data can impede STLS's ability to select the most appropriate model.
STLS' modest performance in noisy and scarce data is partly due to the
fact that it does not account for the uncertainty in the data, even though this
uncertainty can be a crucial factor to consider when choosing a model.
Furthermore, the performance of STLS deteriorates when the number of
data points is lower than a certain threshold. 
As we will demonstrate later in \S\ref{subsec:scaling}, by adopting a Bayesian viewpoint, 
one can better find the reason for such deterioration and the remedy against it, 
effectively addressing the constraints posed by scarce and noisy data.

In the Bayesian point of view, in contrast to STLS, 
models are selected not by the
magnitude of the coefficients of each term but by the
marginalised likelihood (i.e.~evidence), which is proportional to the
probability of the selected model given the data. This procedure of
selecting the model that maximises the evidence is also known as type-II
maximum likelihood and is arguably a better method to evaluate how
suitable the model is to describe the data than STLS. Firstly, this
method naturally avoids overfitting and promotes sparsity, as the
evidence naturally includes an `Occam factor' that penalises
over-parameterised models \cite{Mackay1992}. Compared to the use of the
Pareto front or information criterion to tune hyperparameters, the
Bayesian approach provides a more natural way to determine the parsimony
of the model according to the level of uncertainty in the data. 
The Bayesian account of uncertainty in the data also makes it more resilient against large noise.
Secondly, model selection in this
Bayesian framework is agnostic to the magnitude of the coefficients. As
a result, it can effectively acquire terms with small coefficients,
which is challenging for STLS. Finally, it is well known that many
sparse regression techniques and information criteria can be
re-interpreted from a Bayesian perspective
\cite{Poggio1985, Titterington1985}. For instance, LASSO can be
considered as the maximum a posteriori estimate under a Laplace prior,
whereas thresholding is equivalent to empirically removing terms that
have little significance to the likelihood. The BIC is rooted in the
Bayesian framework and can be regarded as an asymptotic estimate of the
marginalised likelihood at the high data limit, whereas the AIC can be
seen as an extension of the BIC with a different prior
\cite{Burnham2004}. 

Therefore, the goal of this study is to demonstrate how a Bayesian perspective can encompass all preceding techniques
and is best suited for extending the current SINDy techniques to
function in challenging data environments where data are noisy and scarce.
As we shall demonstrate later in \S\ref{subsec:scaling}, by grounding SINDy in the Bayesian framework, we can also more rigorously justify how and why SINDy works well in the high data low noise limit.

\hypertarget{sec:BayesianWork}{%
\subsection{Related work and contribution of this
work}\label{sec:BayesianWork}}

Sparse Bayesian learning (or regression) is a mature and well-established
method in the statistical literature \cite{Mackay1992, Mackay2003, Faul2020}. The dictionary learning technique
typically employs priors that promote sparsity. The
current `gold standard' sparsity-inducing priors are the
spike-and-slab-typed priors, but earlier work has also investigated the
use of hierarchical priors. For example, when one combines a Gaussian
prior with variance \(\alpha^{-1}\) with a maximum-entropy hyperprior
on \(\alpha\) (i.e.~a scale uniform distribution), the result is a
sparsity-inducing student-t true prior \cite{Tipping2004}. 
A notable early study that utilises this hierarchical prior to promote sparsity is the
\texttt{RVM/SparseBayes} package developed by Tippings \& Faul
\cite{Tipping2003}. The Relevance Vector Machine (RVM) is a
type-II-maximum-likelihood algorithm that was initially created as a
probabilistic alternative to the Support Vector Machine (SVM). 
It fits the data to a dictionary made up of kernels similar to those in SVM, it does so by maximising evidence rather than likelihood, which favours sparsity.
This method exploits the
fact that the conditional hyperprior given the data can be approximated
as a Dirac Delta, while the rest of the probability distributions can
be approximated as Gaussians, which allows for analytical
tractability. The resulting approach has been demonstrated to be both
computationally efficient and robust, not only for classification tasks
but also for addressing general sparse regression problems. Due to its
Bayesian foundation, it is also resilient to noisy and scarce datasets.
Therefore, it is especially appropriate for the objective of this study.

We note that this study is not the first to apply sparse Bayesian
regression in the context of equation discovery or system identification
\cite{Hong2008, Pan2016, Niven2024}, nor the first to integrate sparse Bayesian
regression with SINDy \cite{Zhang2018, Fuentes2021}. For instance,
Zhang \& Lin \cite{Zhang2018} incorporated an additional thresholding
procedure into the \texttt{SparseBayes} method and applied the technique
to extract different ordinary differential equations (ODEs) and partial
differential equations (PDEs) from numerical data. They conducted a
comparative analysis of their threshold sparse Bayesian regression
algorithm with STLS and LASSO, and demonstrated Bayesian regression's
superior performance in the presence of noisy data. 
Recent work by Niven \emph{et al.} \cite{Niven2024} has also attempted to use Bayesian inference for model selection and uncertainty quantification in system identification.
However, the synthetic dataset they considered only introduced uncertainty in the time derivatives and not in the state variables themselves, which can oversimplify the challenge of the system identification problem, as we shall demonstrate later in (\ref{eq:total_noise}).
Fuentes \emph{et
al.} \cite{Fuentes2021} substituted STLS with \texttt{SparseBayes}
directly as the sparse regression algorithm and employed the technique
for system identification. In particular, apart from synthetic data,
they have successfully applied this approach to two experimental benchmarks.
Nevertheless, the dataset they used has a high sampling rate (i.e.~a high data limit). 
When finite difference was applied, the noise was
significantly amplified in their time derivative, resulting in
excessively large uncertainty in their learning result despite
successfully recovering the desired equations.

While the above work uses variants of \texttt{SparseBayes}, which do not
require expensive computation, they rely on Gaussian approximations of
the likelihood and prior distribution in order to maintain the
analytical tractability of the distributions. In contrast, subsequent
work \cite{Nayek2021, Martina-Perez2021, Hirsh2022, Wang2022} use
spike-and-slab-typed priors, which, despite the more accurate recovery
of the correct model, necessitates costly sampling techniques, such as
MCMC \cite{Nayek2021, Hirsh2022} or
Approximated Bayesian Computation \cite{Martina-Perez2021} to capture
the distribution of the estimated parameter more accurately. The benefit
of using these sampling-based techniques lies in their capacity to
precisely capture non-Gaussian distributions in the parameter and their
ability to integrate sparsity-promoting spike-and-slab-typed prior.
However, they come at the expense of expensive computations.
A cheaper alternative is to make use of ensembling technique \cite{Fasel2022} to estimate uncertainty in the model. Recent work \cite{Gao2023} has demonstrated the convergence of the bootstrapping-based STLS algorithm to the equivalent Bayesian statistics. 
Nonetheless, these algorithms remain an order of magnitude more expensive than STLS and the Gaussian approximations used by \texttt{SparseBayes}.

Given that our goal is to seek a fast and computationally inexpensive
learning method, this study will follow the earlier work more closely
\cite{Zhang2018, Fuentes2021} and shall adopt strategies similar to the
\texttt{SparseBayes} method. Despite its similarity with the earlier
work, we note a few key differences. First, our work highlights the
key strength of the Bayesian framework, namely, its superior performance
in scarce and noisy data. This is particularly in contrast to
\cite{Fuentes2021}, whose dataset is in the high data limit. Second,
our work compares the performance of the weak formulation and finite
difference in obtaining derivatives from scarce and noisy data and
highlights the pros and cons of each method, which is in contrast to the
finite difference used in earlier work \cite{Zhang2018, Fuentes2021}.
Lastly, we shall present a new sparse Bayesian regression algorithm,
which, despite heavy inspiration from \texttt{SparseBayes}, gives better
performance due to its ability to keep track of how noise variance
propagates throughout the calculation.

\hypertarget{structure-of-the-paper}{%
\subsection{Structure of the paper}\label{structure-of-the-paper}}
The rest of the paper is structured as follows. 
In \S\ref{sec:method}, we will provide a concise overview of the SINDy framework. 
Then, we will introduce our Bayesian approach and estimation used to approximate how noise propagate through the collection of potential functions and temporal derivatives. 
The limitation of these approximations will be discussed later in \S\ref{subsec:limitation}.
We will also briefly compare our Bayesian algorithm with previous studies, and highlight the potential bias introduced by the weak formulations when the sampling rate is low, which further reinforces our case for a noise-robust algorithm (\S\ref{subsec:FastBayes}). 
In \S\ref{sec:examples}, we will showcase the efficacy of the suggested framework by presenting examples such as the Van Der Pol system, the nonlinear oscillator, the Lorenz system, and predator-prey dynamics. 
The predator-prey dynamics are based on real-life data, 
highlighting the practicality of the framework in real-world applications. 
Lastly, we will analyse the reasons why the Bayesian approach is more resilient in scarce and noisy data compared to STLS (\S\ref{subsec:scaling}), 
but not as resilient as traditional Bayesian techniques due to the trade-offs made to speed up computation (\S\ref{subsec:limitation}). 
We will also discuss how the improved data efficiency and the probability framework create new potential innovations in active learning  (\S\ref{subsec:Active}).

\section{Methodology}\label{sec:method}
\subsection{SINDy and tracing noise propagation}\label{sec:SINDy}

Suppose we have a system of interest with state variables
$\hat{\mathbf{x}}$ with dimension \(D\), defined as $\hat{\mathbf{x}}(t) := [\hat{x}_1(t), \; \hat{x}_2(t) \; ... \;\hat{x}_D(t)] \in \mathbb{R}^{1 \times D}$, where $t$ is time. The dynamics of $\hat{\mathbf{x}}(t)$ is governed by the dynamical equation
\begin{equation}
  \dot{\hat{\mathbf{x}}} = \frac{d}{d t}\hat{\mathbf{x}} = f(\hat{\mathbf{x}})
\label{eq:dyn_clean}\end{equation} with an unknown function
\(f: \mathbb{R}^{D} \rightarrow \mathbb{R}^{D}\). 
Suppose the system was
observed $\tilde{N}$ times at \(t=t_1,\; t_2,\;...t_{\tilde{N}}\) and the state variables
are measured, resulting in the dataset \begin{equation}
\mathbf{X}=
\begin{bmatrix}
\mathbf x_1 & \mathbf x_2 & ... & \mathbf x_D
\end{bmatrix}
= 
\begin{bmatrix}
 x_1(t_1) &  x_2(t_1) & ... & x_D(t_1)\\
 x_1(t_2) &  x_2(t_2) & ... & x_D(t_2)\\
\vdots & \vdots & \ddots &\vdots\\
 x_1(t_{\tilde{N}}) &  x_2(t_{\tilde{N}}) & ... & x_D(t_{\tilde{N}})\\
\end{bmatrix}
\in \mathbb{R}^{\tilde{N} \times D}.
\label{eq:XDiscrete}\end{equation} 
Due to measurement error, the data might be contaminated by noise
\(\boldsymbol{\epsilon}_x\), such that the observed variables are: 
\begin{equation}
X_{ij} = x_j(t_i) = \hat{x}_{j}(t_i)+\epsilon_{x,ij}.
\label{eq:X_AddNoise}\end{equation} 
Here, we assume that
the noise
\({\boldsymbol{\epsilon}}_x \in \mathbb{R}^{\tilde{N} \times D}\)
consists of \(\tilde{N} \times D\) independent random variables with Gaussian
distributions, with zero means and variance
\(\boldsymbol\sigma_x^2 \in \mathbb{R}^{\tilde{N} \times D}\).
This simplifiying assumption is introduced mostly for pedagogical simplicity.
More discussion on potential extension to the noise model beyond independent Gaussian noise will be give in \S\ref{subsec:limitation}.

The goal of SINDy is to discover \(f\) from the observed time series
data \(\mathbf{X}\). SINDy postulates that \(f\) can be written as a
linear combination of a library of candidate functions. For example, a
commonly used library is the polynomial library of the state variables: \[
\boldsymbol{\Theta}(\hat{\mathbf{x}}) = \begin{bmatrix}{1} & \hat{x}_{1} & \hat{x}_{2} & \hat{x}_{1}^2 & \hat{x}_{1} \hat{x}_{2} & \hat{x}_{2}^2 & \hat{x}_{1}^3 & ...\end{bmatrix},
\] or when applied to the observed dataset, \[
\boldsymbol{\Theta}(\mathbf{X}) = 
\begin{bmatrix}
\mathbf{1} & \mathbf{x}_1 & \mathbf{x}_2 & \mathbf{x}_1^2 & \mathbf{x}_1 \mathbf{x}_2 & \mathbf{x}_1^2 &\mathbf{x}^3 & ...
\end{bmatrix}
\in \mathbb{R}^{\tilde{N} \times M},
\] in which we assume the library consists of \(M\) columns of possible candidates. Here, the product and power operations
\(\mathbf{x}^n \mathbf{y}^m = [x_{i}^n y_{i}^m]\) are applied
elementwise.
If \(f\) can be described by a few terms in the library of candidate
functions, then the parameter vector \(\mathbf{w}\) that fulfils
\begin{equation}
\dot{\hat{\mathbf{x}}}_c =\boldsymbol{\Theta}(\hat{\mathbf{x}}) \mathbf{w}
\label{eq:dyn_clean_w}\end{equation} will be sparse. The aim of SINDy is
to recover this sparse parameter vector \(\mathbf{w}\) through
regression. 
If, however, \(f\) cannot be fully described by a linear composition
of library candidate functions, but can be approximated by it, then
SINDy finds a parsimonious approximation to \(f\) through sparse
regression.

In practice, the time derivative is usually not measured as
part of the dataset. To recover the time derivative from the time-series
data \(\mathbf{X}\), we assume there exist some linear operators \(L_{\partial t}\) and \(L_I\)
that approximates the time derivative and the interpolation at $N$ time points, respectively, such that
\begin{equation}
L_{I} \dot{\mathbf{X}} \approx L_{\partial t} \mathbf{X}, \quad \mbox{and therefore}  \quad L_{\partial t} \mathbf{X} \approx  L_{I} f(\mathbf{X})
\label{eq:Discrete-Diff}\end{equation} in the absence of noise,
i.e.~when \(\boldsymbol{\epsilon}_x=0\). 
Therefore, \(L_I\) and
\(L_{\partial t}\) are matrices with dimensions \(N \times \tilde{N} \). 
Further discussion on these operators for recovering time derivatives will follow in \S\ref{sec:WeakVSFD}.

Substituting (\ref{eq:X_AddNoise}) and (\ref{eq:Discrete-Diff}) into (\ref{eq:dyn_clean_w}), we rewrite the regression as 
\begin{equation}
L_{\partial t} \mathbf{X} = L_I\boldsymbol{\Theta}(\mathbf{X})\mathbf{w} + \boldsymbol{\epsilon}.
\label{eq:dyn_dis_noise}\end{equation} 
The total noise \(\boldsymbol{\epsilon}\) originates from the propagation of \(\boldsymbol{\epsilon}_x\) through the transformations in (\ref{eq:dyn_dis_noise}).
It is probably analytically intractable to calculate the full distribution of \(\boldsymbol{\epsilon}\), which motivates the use of sampling methods like MCMC.
Nevertheless, the variance of \(\boldsymbol{\epsilon}\) can be estimated by tracing how the noise variance propagates through (\ref{eq:dyn_dis_noise}), which can then be used to guide model selection under the Bayesian framework.
Although this variance approximation is crude, it allows for inexpensive computation, which is particularly crucial in real-time applications. 
In other words, this study foregoes a complete and accurate calculation of the noise distribution in favour of a rapid approximation that enables real-time model selection or discovery.
Further analysis of the constraints of this approximation will follow in
 \S\ref{subsec:limitation}.

Here, we outline our method for estimating the variance $\boldsymbol\sigma^2$ of the total noise $\boldsymbol{\epsilon}$.
Firstly, we assume that the statistics of noise arising from each nonlinear function (each column) in \(\boldsymbol{\Theta}(\mathbf{X})\) and time derivative \(L_{\partial t}{\mathbf{X}}\) are independent.
Secondly, we assume that the noise at each time point in (\ref{eq:dyn_dis_noise}) is independent of each others, i.e. we approximate the covariance matrix by its diagonal.
These approximations circumvent the calculation of correlations between nonlinear terms, which is combinatorially expensive, and between time points, which necessitates the storage and inversion of a large covariance matrix.
These are \emph{ad hoc} approximations aimed at reducing the computational cost, and their theoretical justification may require further investigation.
However, we can empirically show their effectiveness in the examples later (e.g. Figure \ref{fig:OscVar}).

Because each term is assumed to be independent, we can break down the computation of the noise variance of the total noise \(\boldsymbol{\epsilon}\) into contributions from each term in (\ref{eq:dyn_dis_noise}).
The contribution from the time derivative \(\boldsymbol{\sigma}_{\partial t}\) is given by $\boldsymbol\sigma^2_{\partial t} = L_{\partial t}^2 \boldsymbol\sigma_{x}^2 $, assuming \(L_{\partial t}^2 \) is the elementwise square of the matrix representing the linear operator $L_{\partial t}$. 
(Note that the full covariance matrix should be $L_{\partial t} \mbox{diag}(\boldsymbol\sigma_{x}^2) L_{\partial t}^T$ for each column of $\boldsymbol\sigma_{x}^2$. However, since the variance in the diagonal dominates over the off-diagonal terms, we assume temporal independence and neglect the off-diagonal contribution to reduce computational cost.) 
The contribution from the library can be estimated by a Taylor expansion of statistical moments \cite{Wolter1985}. 
Since nonlinear functions are applied elementwise, we can construct a corresponding library of variance, 
\begin{equation}
\boldsymbol\sigma^2_\Theta(\mathbf{x},\boldsymbol\sigma_x^2) = 
\begin{bmatrix}
\boldsymbol\sigma^2_1 & \boldsymbol\sigma^2_{x_1} & \boldsymbol\sigma^2_{x_2} & \boldsymbol\sigma^2_{x^2_1} & \boldsymbol\sigma^2_{x_1 x_2} & \boldsymbol\sigma^2_{x^2_2} & \boldsymbol\sigma^2_{x^1_3} & ...
\end{bmatrix}
\label{eq:lib_noise}\end{equation} where each element is the variance of the corresponding element in $\boldsymbol{\Theta}(\mathbf{X})$.
The exact formula used to calculate the variance of each term in the library is given in Appendix A.
By the above approximations, the variance $\boldsymbol{\sigma}^2$ of the total noise
\(\boldsymbol{\epsilon}\) can be written as \begin{equation}
\boldsymbol{\sigma}^2 = L_I^2\boldsymbol{\sigma}^2_\Theta(\mathbf{x},\boldsymbol{\sigma}_x^2) \mathbf{w}^2 + L_{\partial t}^2 \boldsymbol\sigma_x^2,
\label{eq:total_noise}\end{equation} where \(\mathbf{w}^2 = [w_{ij}^2]\)
is the elementwise square of \(\mathbf{w}\) and $L_I^2$ the elementwise square of $L_I$.
Notice that the noise variance $\boldsymbol\sigma^2$ depends on the parameter \(\mathbf{w}\). 
This dependency was omitted in previous work \cite{Zhang2018,Niven2024,Fuentes2021},
but this study will address this dependency in the regression step.

For simpler notation, we define the left-hand side of the regression in
(\ref{eq:dyn_dis_noise}) as \begin{equation}
\mathbf{y} = L_{\partial t} \mathbf{X}
\in \mathbb{R}^{N \times D}
\label{eq:t_def}\end{equation} and the dictionary, which is the library
of functions of the measured state variables \(\mathbf{X}\), as
\begin{equation}
\mathbf{D}(\mathbf{X}) = L_I\boldsymbol{\Theta}(\mathbf{X}) = 
\begin{bmatrix}
\mathbf{d}_1 & \mathbf{d}_2 & \mathbf{d}_3 & ... & \mathbf{d}_m & ... & \mathbf{d}_M
\end{bmatrix}
\in \mathbb{R}^{N \times M}
\label{eq:D_def}\end{equation}

Substituting the redefined variables \(\mathbf{y}\) and \(\mathbf{D}\)
into (\ref{eq:dyn_dis_noise}), the sparse regression problem becomes
\begin{equation}
\mathbf{y} = \mathbf{D}\mathbf{w} + \boldsymbol{\epsilon}.
\label{eq:tDw}\end{equation} 
As a final approximation, we approximate
\(\boldsymbol{\epsilon} \sim \mathcal{N}(0,\boldsymbol\beta^{-1})\) as a Gaussian noise with zero mean and elementwise
variance \(\boldsymbol\beta^{-1} = \boldsymbol\sigma^2\). Here, we use
the inverse noise variance \(\boldsymbol\beta^{-1}\) instead of
\(\boldsymbol\sigma^2\) to facilitate easier notation in the next
section.
We assume this so that we can employ the following fast Bayesian regression algorithm. 
The validity of this approximation will be discussed in \S\ref{subsec:limitation}.

\subsection{Fast Bayesian Regression under the Gaussian
assumption}\label{subsec:FastBayes}

In some fields, such as control engineering, the Bayesian approach is avoided because accurate calculations of the statistical distributions are analytically intractable, 
forcing one to resort to expensive Monte Carlo simulations or other sampling techniques. 
Nonetheless, if we accept the Gaussian approximation for the likelihood as discussed in (\ref{eq:tDw}) and combine it with a Gaussian prior, we can maintain analytical tractability throughout the rest of the calculation. 
Furthermore, for model discovery, a zero-mean Gaussian prior is a good proxy for an objective prior. 
In theory, an objective prior for the parameter \(\mathbf{w}\) can be defined hierarchically, in which the prior on the parameter $\mathbf{w}$ is given by the zero-mean Gaussian $p(\mathbf{w}|\boldsymbol\alpha) = \mathcal{N}(0,\boldsymbol\alpha^{-1})$, and a hyperprior on the inverse variance $\boldsymbol\alpha^{-1}$ is given by a log-uniform distribution. 
The hyperprior ensures that the overall prior contains no information on the scale of $\mathbf{w}$. 
However, previous work \cite{Mackay1999} has shown that such a hierarchical prior may significantly bias the maximum a posteriori estimate of the parameter \(\mathbf{w}\). 
Instead, optimising the hyperparameter \(\boldsymbol\alpha\) to maximise the evidence can provide a good estimate of integrated evidence while avoiding the bias introduced by the hierarchical prior. 
In other words, a zero-mean Gaussian prior with an optimal hyperparameter \(\boldsymbol\alpha\) is a good approximate objective prior for sparse regression \cite{Tipping2004}.
It is on this premise that \texttt{SparseBayes} was developed, while this study will also follow the same principle.

In the following section, we will introduce the Bayesian view of sparse
regression and derive our regression method alongside
\texttt{SparseBayes}. For pedagogical reasons, in this section we will
focus on the regression of a single state variable, i.e.~a single column
in \(\mathbf{y}\) (and \(\mathbf{w}\)). One can easily extend the method
to multiple dimensions later by repeating the regression for each state
variable.

In the Bayesian framework,
instead of seeking \(\mathbf{w}\) that best fits (\ref{eq:tDw}), one
seeks the posterior distribution of \(\mathbf{w}\) conditioned on the
dataset \((\mathbf{D}, \mathbf{y})\), which can be calculated using
Bayes' rule \begin{equation}
p(\mathbf{w} | \mathbf{y}, \mathbf{D}, \boldsymbol\beta,\boldsymbol\alpha) = 
\frac{p(\mathbf{y}|\mathbf{D},\mathbf{w},\boldsymbol\beta) 
p(\mathbf{w}|\boldsymbol\alpha)}
{p(\mathbf{y}|\mathbf{D},\boldsymbol\beta,\boldsymbol\alpha)}.
\label{eq:BayesRule}\end{equation} Here,
\(p(\mathbf{y}|\mathbf{D},\mathbf{w},\boldsymbol\beta)\) is the
likelihood and \(p(\mathbf{w}|\boldsymbol\alpha)\) is the prior
distribution with the hyperparameter
\(\boldsymbol\alpha \in \mathbb{R}^M\) , elements of which define the
inverse of the prior variance of each element in \(\mathbf{w}\). The
evidence
\(p(\mathbf{y}|\mathbf{D},\boldsymbol\beta,\boldsymbol\alpha) = \int p(\mathbf{y}|\mathbf{D},\mathbf{w},\boldsymbol\beta) p(\mathbf{w}|\boldsymbol\alpha) d \mathbf{w}\)
normalises the posterior distribution. If we assume that the prior is
Gaussian with variance \(\boldsymbol\alpha^{-1}\) ,
i.e.~\(p(\mathbf{w}|\boldsymbol\alpha) \sim \mathcal{N}(\mathbf{0},\boldsymbol\alpha^{-1})\),
and the likelihood can be approximated as Gaussian under the assumption \(\mathbf{\epsilon} \sim \mathcal{N}(0,\boldsymbol\beta^{-1})\), then
all of the above distributions are Gaussians and can be written analytically. 
In particular, the log-evidence can be written as
\cite{Tipping2003} \begin{equation}
\mathcal{J}(\mathbf{y}|\mathbf{D},\boldsymbol\beta,\boldsymbol\alpha) = -\frac{1}{2}(\ln|{2\pi} \mathbf{C}|+\mathbf{y}^T\mathbf{C}^{-1}  \mathbf{y}).
\label{eq:log-evidence}\end{equation} where \begin{equation}
\mathbf{C} = \mathbf{B}^{-1} + \mathbf{D}\mathbf{A}^{-1}\mathbf{D}^T = \mathbf{B}^{-1} + \Sigma_{m=1}^M \frac{1}{\alpha_m} \mathbf{d}_m \mathbf{d}_m^T  \in \mathbb{R}^{N \times N},
\label{eq:C_def}\end{equation} where \(\mathbf{A}\) is a diagonal matrix
with entries \(A_{mm}=\alpha_m\), \(\mathbf{B}\) is a diagonal matrix
with entries \(B_{nn}=\beta_n\), and \(|\mathbf{C}|\) denotes the
determinant.

Now, during regression, one can perform model selection by selecting the
columns \(\mathbf{d}_m\) in \(\mathbf{D}\) that should be included in
the model while removing columns that should not be in the model. In
STLS, columns are removed by setting the corresponding \(w_m\) to zero
if it is smaller than a certain threshold. In a Bayesian framework,
models are compared by their evidence, so we seek to maximise the
log-evidence
\(\mathcal{J}(\mathbf{y}|\mathbf{D},\boldsymbol\beta,\boldsymbol\alpha)\)
by removing columns from \(\mathbf{D}\) . In the calculation in
(\ref{eq:C_def}), this is equivalent to setting the corresponding
elements in \(\boldsymbol\alpha\) to infinity. 
Now, it is interesting to
isolate how each element \(\alpha_m\) and the corresponding column
\(\mathbf{d}_m\) changes
\(\mathcal{J}(\mathbf{y}|\mathbf{D},\boldsymbol\beta,\boldsymbol\alpha)\).
By defining \begin{equation}
\mathbf{C}_{-m} = \mathbf{C} - \frac{1}{\alpha_m} \mathbf{d}_m \mathbf{d}_m^T,
\label{eq:C-m_def}\end{equation} and using the Sherman-Morrison formula,
we isolate the contribution of the \(m\)-th term
\(\mathcal{J}(\alpha_{m})\) from the evidence without the \(m\)-th term
\(\mathcal{J}(\boldsymbol\alpha_{-m})\),
i.e.~\(\mathcal{J}(\mathbf{y}|\mathbf{D},\boldsymbol\beta,\boldsymbol\alpha) = \mathcal{J}(\boldsymbol\alpha_{-m}) +\mathcal{J}(\alpha_{m})\),
and find that \begin{equation}
\mathcal{J}(\alpha_{m}) = \frac{1}{2} \left( \ln{\alpha_m} - \ln{(\alpha_m + \mathbf{d}_m^T \mathbf{C}^{-1}_{-m}\mathbf{d}_m) + \frac{(\mathbf{d}_m^T \mathbf{C}^{-1}_{-m}\mathbf{y})^2}{\alpha_m+\mathbf{d}_m^T \mathbf{C}^{-1}_{-m}\mathbf{d}_m}} \right).
\label{eq:J_ind}\end{equation}

\begin{figure}
\hypertarget{fig:Jalpha}{%
\centering
\includesvg[ width = 1 \columnwidth ]{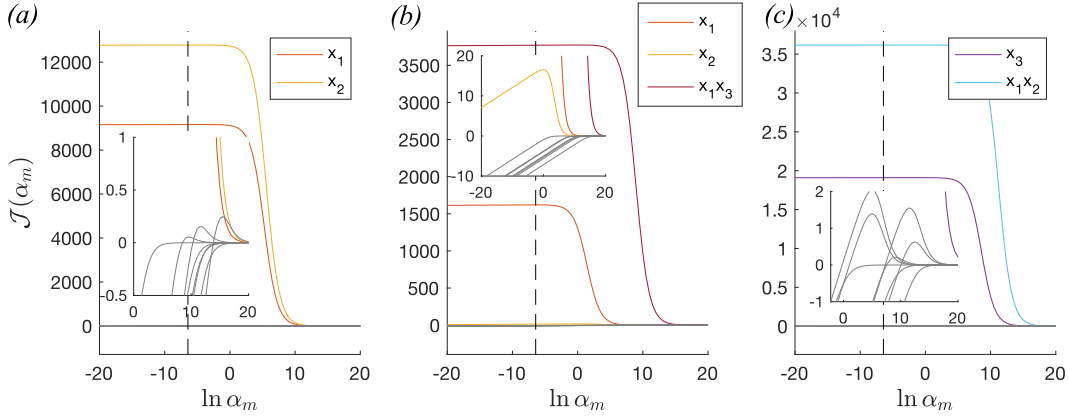}
\caption{Contribution of the \(m\)-th term in the model to the evidence
at given prior variance \(\alpha_m\). Data are generated from a
Lorenz system, with noise \(\sigma_x=0.27\) added to 201 data points.
Terms in the model are highlighted by their respective colours, whereas
terms that should not be in the model are grey. The inset shows the
zoomed-in picture of the same graph near \(\mathcal{J}(\alpha_{m})=0\).
Notice that terms that should be included have dramatically positive
\(\mathcal{J}(\alpha_{m})\), except for \(x_2\) in \((b)\), which has a
relatively small but positive contribution. Also, in the inset of
\((a,c)\), some grey lines are slightly above zero in small regions of
\(\alpha_m\), but remain negative when \(\alpha_m\) is small enough. For
our method, \(\alpha_m\) is fixed at a small value, as shown by the
black dashed line.}\label{fig:Jalpha}
}
\end{figure}

Figure \ref{fig:Jalpha} shows the typical \(\mathcal{J}(\alpha_{m})\)
for the correct terms that should be included in a Lorenz system (colored) and the incorrect terms that should
not be in the equations (c.f. (\ref{eq:Lorenz})) 
when the algorithm is trying to learn them from noisy data of the system. 
Asymptotically, \(\mathcal{J}(\alpha_{m})\) always approaches zero
when \(\alpha_m \rightarrow \infty\), because it is equivalent to not
including the \(m\)-th term in the model. Meanwhile, as
\(\alpha_m \rightarrow -\infty\), \(\mathcal{J}(\alpha_{m})\) approaches
negative infinity, which is equivalent to setting prior and evidence vanishingly small. 
Note that, for all incorrect terms, the corresponding
\(\mathcal{J}(\alpha_{m}) \leq 0\) for all \(\alpha_m \in \mathbb{R}\),
implying that the inclusion of the term will always decrease evidence.
Meanwhile, for the correct terms,
\(\mathcal{J}(\alpha_{m})>0\) in a large range of \(\alpha_m\) and \(\mathcal{J}(\alpha_{m})\)
inevitably exhibits a peak at
\begin{equation}
\alpha_m = \frac{(\mathbf{d}_m^T \mathbf{C}^{-1}_{-m} \mathbf{d}_m)^2}{(\mathbf{d}_m^T \mathbf{C}^{-1}_{-m}\mathbf{y})^2-\mathbf{d}_m^T \mathbf{C}^{-1}_{-m} \mathbf{d}_m}.
\label{eq:opt_alpha}
\end{equation}

Therefore, one can determine if a term should be included in the model
by observing if \(\mathcal{J}(\alpha_{m})>0\) for some \(\alpha_m\),
which can be determined by
\((\mathbf{d}_m^T \mathbf{C}^{-1}_{-m}\mathbf{y})^2-\mathbf{d}_m^T \mathbf{C}^{-1}_{-m} \mathbf{d}_m>0\).
However, it should be noted that \(\mathcal{J}(\alpha_{m})\) depends on
\(\mathbf{C}_{-m}\), which changes according to the existence of other
terms in the model. As other columns \(\mathbf{d}_{i\neq m}\) are added
or subtracted from the model, \(\mathcal{J}(\alpha_{m})\) for the column
\(\mathbf{d}_{m}\) will need to be recomputed.

The \texttt{SparseBayes} algorithm exploits this observation and devises an
iterative algorithm that optimises each element in \(\boldsymbol\alpha\) to
maximise \(\mathcal{J}(\alpha_{m})\), and removes incorrect terms that do not
contribute to evidence positively, thereby promoting sparsity. Another
interpretation is that this algorirthm effectively
makes the Gaussian prior a sparsity-promoting spike-and-slab prior.
Since the algorithm always seeks to maximise \(\mathcal{J}(\alpha_{m})\)
for each \(\alpha_m\), incorrect terms will have the corresponding \(\alpha_m\) optimised to 
\(\alpha_m \rightarrow \infty\), effectively making the prior a spike at
zero, while terms that should be included in the model will have a
finite \(\alpha_m\) value, which is effectively a slab prior. As
for \(\boldsymbol\beta\), the \texttt{SparseBayes} algorithm accepts a
prescribed value \(\sigma^2\) for all data points,
i.e.~\(\mathbf{B} = \sigma^{-2}\mathbf{I}\). Alternatively, the
algorithm can optimise a single \(\sigma^2\) value for maximum evidence, but does not accept \(\sigma^2\) varying across different time points.

Here, we present a novel Bayesian regression approach that also
maximises evidence, but differs from \texttt{SparseBayes} in the
optimisation of \(\boldsymbol\alpha\) and \(\boldsymbol\beta\). Phenomenologically, we noticed that in some cases, incorrect terms have
\(\mathcal{J}(\alpha_{m})\) values exceeding zero in some small regions
of \(\alpha\), which might be attributed to incidental matching between
spurious terms and noise.
\texttt{SparseBayes} often mistakenly captures these spurious terms, usually with small coefficients. 
Zhang \& Lin
\cite{Zhang2018} addressed this problem by incorporating a thresholding
procedure following the regression performed by the \texttt{SparseBayes}
algorithm. In contrast, this study introduces a technique that
circumvents the optimisation process of \(\boldsymbol\alpha\) entirely.
Instead, based on the observation that \(\mathcal{J}(\alpha_{m})\) for the correct terms is always positive for a large range of \(\alpha_m\), we set \(\boldsymbol\alpha\) to a certain low value, ensuring
that \(\mathcal{J}(\alpha_{m})\) remains positive for the correct
terms, while becoming negative for the incorrect ones. 
The value of \(\boldsymbol\alpha\) can often be approximated by expert prior knowledge or physical scaling arguments, and it represents our prior knowledge of the system.
We found that often the value of \(\boldsymbol\alpha\) is not critical, as long as it is small enough to ensure that \(\mathcal{J}(\alpha_{m})\) remains positive for the correct terms.
In another viewpoint,
\(\boldsymbol\alpha\) is effectively a tunable hyperparameter in our
proposed method, and it represents the a weakly informative prior on the scale of the parameter \(\mathbf{w}\).

\begin{algorithm}[H]
  \SetKwInOut{Input}{input}
  \SetKwInOut{Output}{output}
  \Input{Library $\mathbf{D}(\mathbf{X})$, dynamics $\mathbf{y}$, library noise variance $L_I^2 \boldsymbol\sigma_\Theta^2$, dynamics noise variance $L_{\partial t}^2 \boldsymbol\sigma_{x}^2$, prior variance inverse $\boldsymbol\alpha$}
  \Output{Learnt parameter $\mathbf{w}$}
  initialization\;
  \Repeat{further removal of columns does not increase evidence}{
    \For{$i$th column in $\mathbf{D}$}
    { Set $\mathbf{\tilde{D}}$ that includes all columns in $\mathbf{D}$ except the current $i$th column\;
      Perform ridge regression $\min_\mathbf{w} (\tfrac{1}{2}||\boldsymbol\beta(\mathbf{y}-\mathbf{D}\mathbf{w})||^2_2 + \tfrac{1}{2}||\boldsymbol\alpha \mathbf{w}||^2_2)$ to get $\mathbf{w}$, using $\boldsymbol\beta= (L_{\partial t}^2 \boldsymbol\sigma_{x}^2)^{-1}$ as the inverse noise variance and the prescribed $\boldsymbol\alpha$ as inverse prior variance\;
      \While{changes in $\mathbf{w}$ is greater than a threshold}
      { Update $\boldsymbol\beta$ using (\ref{eq:total_noise}), with $\mathbf{w}$ from the previous iteration\;
        Perform ridge regression $\min_\mathbf{w} (\tfrac{1}{2}||\boldsymbol\beta(\mathbf{y}-\mathbf{D} \mathbf{w})||^2_2 + \tfrac{1}{2}||\boldsymbol\alpha \mathbf{w}||^2_2)$ to update $\mathbf{w}$, using the updated $\boldsymbol\beta$ as the inverse noise variance and the prescribed $\boldsymbol\alpha$ as inverse prior variance\;}
      Calculate and store the evidence $\mathcal{J}(\mathbf{y}|\mathbf{\tilde{D}} ,\boldsymbol\beta,\boldsymbol\alpha)$ using (\ref{eq:log-evidence})\;
    }
    Select the column that yields the greatest evidence after removal, and remove this column in $\mathbf{D}$ \;
  }
  \caption{Greedy algorithm for evidence maximisation (with noise-parameter iteration)}\label{alg:greedy}
\end{algorithm}

Another distinguishing feature of our method compared to \texttt{SparseBayes} and previous studies \cite{Zhang2018,Niven2024,Fuentes2021} is
its treatment of the noise inverse variance \(\boldsymbol{\beta}\).
Previous methods assumed that the noise level is a constant across all
data points, but this assumption may impede the algorithm's ability to
learn the optimal outcome. Instead, (\ref{eq:total_noise}) and (\ref{eq:C_def}) showed a more
precise, 
albeit not the most accurate, approximation to the noise variance, 
in which the variance's dependency on the learnt parameter \(\mathbf{w}\) is taken into account. 
The formulation requires the estimated parameter \(\mathbf{w}\) \emph{a priori}, 
but its value remains unknown until after the regression. 
Hence, we propose an algorithm for an
iterative technique to determine \(\boldsymbol{\beta}\) by considering
the \(\mathbf{w}\)-dependent component of equation
(\ref{eq:total_noise}) as a correction. Initially, we conduct regression
under the assumption that \(\mathbf{w}=0\), which is consistent with the
prior. Following each regression, we modify the value of
\(\boldsymbol\beta = \boldsymbol\sigma^{-2}\) based on the value of
\(\mathbf{w}\) from the last iteration and equation
(\ref{eq:total_noise}). Subsequently, we iterate the regression process
until the modification to \(\boldsymbol{\beta}\) falls below a specific
numerical threshold. Generally, convergence can be achieved within 3-5
iterations.

In order to find the optimal model, specifically the model with maximum
evidence, we suggest using a Greedy algorithm, outlined in algorirthm \ref{alg:greedy}. Starting from the full
model comprising \(M\) terms, we assess the potential increase in
evidence by eliminating each individual term. Subsequently, we choose
\(M-1\) terms that yield the greatest increase in evidence, and iterate
this procedure until any further increase in evidence is impossible. The
technique of our proposed method is outlined in algorithm \ref{alg:greedy}, which
combines our search strategy to maximise evidence with the
aforementioned noise iteration.

The computational complexity of this algorithm scales with the number of
terms in the library \(M\) by \(\mathcal{O}(M^2/2)\). Although this is
slower than the STLS method, it is still much faster than a brute-force
search (\(\mathcal{O}(2^M)\)) and any sampling technique. Furthermore,
our elimination strategy is more robust than the iterative procedure
employed by \texttt{SparseBayes}. In the \texttt{SparseBayes} algorithm,
the search begins with a single term and additional terms are later
added based on the prospective rise in evidence. Although this
technique can be more computationally efficient, its inclusion of both
addition and deletion of terms may result in occasional entrapment in a
loop where collinear terms are repeatedly added and eliminated. Our
elimination technique, in contrast, ensures convergence within \(M^2/2\)
iterations.
It should be noted, however, that the Greedy algorithm does not guarantee convergence to the global maximum, as there is always a chance that the algorithm may prematurely remove a term that should have been included in the model.
Nevertheless, the likelihood of this happening is minimal because terms that were removed early often have a substantial adverse impact on the evidence (see Figure \ref{fig:Jalpha}).
Furthermore, with respect to the convergence of Greedy algorithms, the conventional STLS algorithm employed by the original SINDy framework also exhibits similar features to the proposed algorithm. This is because the thresholding procedure is also essentially a greedy algorithm.  
The issue of convergence to global extrema in STLS was addressed in \cite{Zhang2019}, which may also shed some light on the convergence of the proposed Greedy algorithm.

\subsection{Comparing derivative approximation using weak formulation and finite difference}\label{sec:WeakVSFD}

One of the major challenges in applying SINDy-type algorithms to learn
governing equations from data is approximating temporal (for learning
ODEs) and spatial (for learning PDEs) derivatives using discrete data in
the presence of noise. Conventional finite differences can greatly
amplify noise, rendering the dynamics unrecoverable by the STLS
algorithm. To overcome this problem, one can employ the weak formulation
\cite{Messenger2021, Reinbold2020}. This formulation uses Galerkin
projection and a compact and smooth test function \(\phi(t)\) to
``transfer'' the derivatives on the data to the test function. For
example, the Galerkin projection allows the time derivative on the data
\(\hat{x}^{(k)}(t)\) to be expressed as \[
\int_a^b \hat{x}^{(k)}(t) \phi(t)dt = (-1)^k \int_a^b \hat{x}(t) \phi^{(k)}(t)dt,
\] provided that \(\phi(t)\) and its derivatives up to \(\phi^{(k)}(t)\) are smooth and
have compact support in \(t \in [a,b]\). As a result, any noise present
in the data is not magnified, leading to a substantial enhancement in
the performance of the subsequent regression.

In practice, when applying the method to a discrete dataset
\(\mathbf{X}\), both finite difference and weak formulation can be
written as linear operations on \(\mathbf{X}\), i.e.~as matrices
multiplied to \(\mathbf{X}\), resulting in the formulation in
(\ref{eq:Discrete-Diff}). In fact, both methods result in banded
matrices \begin{equation}
L_{\partial t} =
\begin{bmatrix}
 a_{-n} & a_{-n+1} & ...& a_{n-1} & a_{n} & 0 & ... & 0 \\
 0  & a_{-n} & a_{-n+1} & ...  & a_{n-1} & a_{n} & ... & 0 \\
\vdots& \ddots & \ddots & \ddots & ... & \ddots & \ddots & \vdots \\
0 & ... & 0 & a_{-n} & a_{-n+1} & ...  & a_{n-1} & a_{n}  \\
\end{bmatrix}
\label{eq:LD_Def}\end{equation} and \begin{equation}
L_I =
\begin{bmatrix}
 b_{-n} & b_{-n+1} & ...& b_{n-1} & b_{n} & 0 & ... & 0 \\
 0  & b_{-n} & b_{-n+1} & ...  & b_{n-1} & b_{n} & ... & 0 \\
\vdots& \ddots & \ddots & \ddots & ... & \ddots & \ddots & \vdots \\
0 & ... & 0 & b_{-n} & b_{-n+1} & ...  & b_{n-1} & b_{n}  \\
\end{bmatrix}.
\label{eq:LI_Def}\end{equation} 
For an equispaced central difference method of order \(2n\), \(b_0 = 1\), and
\(b_i = 0, \; \forall i \neq 0\), and \(a_i = \hat{a}_i/\Delta t\),
where \(\Delta t\) is the size of the timestep in the regularly sampled
time-series data, and \(\hat{a}_i\) the coefficients for the central
difference. For example, when \(n=2\),
\([\hat{a}_{-2}\; ... \; \hat{a}_2]=[1/12, \; -2/3, \; 0, \; 2/3, \; -1/12]\).
Meanwhile, the weak formulation gives \[
a_n  = w_{n}\dot\phi_{n} \quad \textrm{and} \quad b_n  = w_{n}\phi_{n}
\] where \(w_{-n}...w_{n}\) gives the quadrature for the integration and
\(\phi_i\) and \(\dot\phi_i\) are the test function and its time
derivative at \(t=(a+b)/2+2i(b-a)/n\).

Note that both \(L_{\partial t}\) and \(L_I\) are rectangular matrices with
dimension \({N} \times \tilde{N} \), where \(\tilde{N}\) is the number of
measurements in the original dataset, whereas \(N \leq \tilde{N}\) is
the number of data after computing the derivatives. \(\tilde{N} -N=2n\)
number of data points are usually filtered away near the two ends of the
data series due to a lack of data beyond the measured point to compute
the corresponding derivative. Also, notice that the scaling of
\(|L_{\partial t}|:|L_I|\) is of order \(\mathcal{O}(1/\Delta t)\) in finite
difference, whereas that of weak formulation is of order
\(\mathcal{O}(1)\). This is why, when applied to
(\ref{eq:Discrete-Diff}), finite difference amplifies noise, while weak
formulation does not.

\begin{figure}
\hypertarget{fig:VanDerPolWeakVSFD}{%
\centering
\includesvg[ width = 1 \columnwidth ]{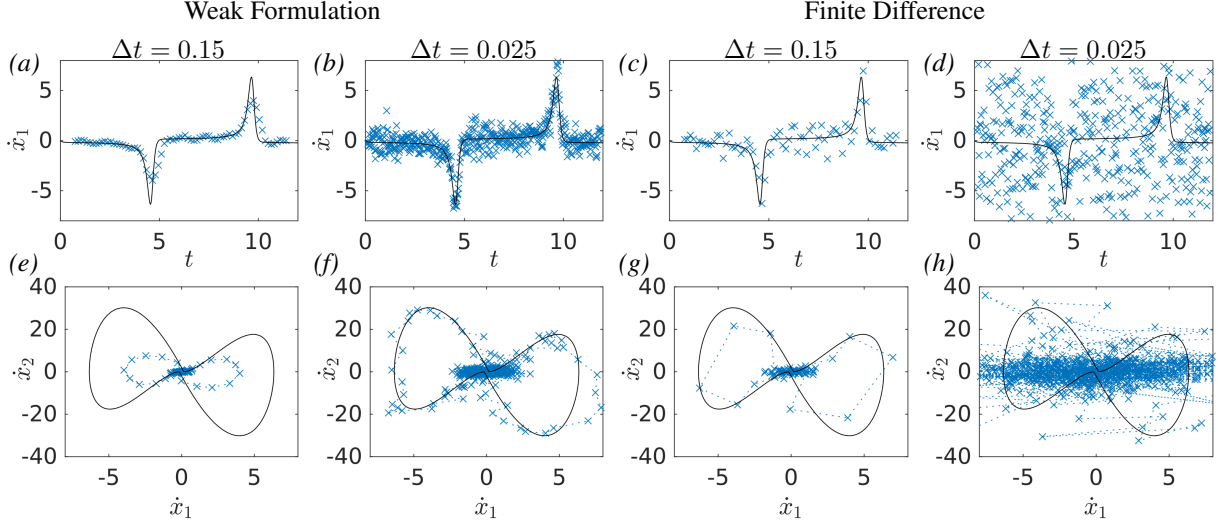}
\caption{Comparison of time derivatives in the Van Der Pol system recovered from noisy data (blue crosses) against the true system (black line), using $(a,b,e,f)$ weak formulation with 9 points and test function $\phi=(t^2-1)^2$ and $(c,d,g,h)$ 8th order (9 points) central difference: $(a$ - $d)$ the time derivative of $x_1$ over time $t$; $(e$ - $h)$ the phase diagram on the $\dot{x}_1-\dot{x}_2$ plane. While the noise $\sigma_x=0.1$ and the number of points used to calculate each derivative $n+1=9$ are fixed, the sampling frequency of the dataset varies between $\Delta t = 0.1$ in $(a,c,e,g)$ and $\Delta t = 0.025$ in $(b,d,f,h)$. Hence, the time window over which the weak formulation is applied varies proportionally to $\Delta t$. Note that in $(a)$, the weak formulation causes significant artificial smoothing due to a large time window, leading to an inaccurate approximation of the phase diagram in $(e)$.}\label{fig:VanDerPolWeakVSFD}
}
\end{figure}

However, from a signal processing standpoint, the weak formulation can
be described as a convolution operator that performs low-pass filtering
\cite{Messenger2021}. This filtering process can remove high-frequency
signals and perhaps result in misidentification of the model. This is
best demonstrated by the phase diagram of the Van der Pol system (figure
\ref{fig:VanDerPolWeakVSFD}), where the dynamic features a sudden and
rapid change. When the sampling rate is high, the weak formulation
effectively captures time derivatives without significantly amplifying
noise (figure \ref{fig:VanDerPolWeakVSFD} \((b,f)\)) compared to the
finite difference method (figure \ref{fig:VanDerPolWeakVSFD} \((d,h)\)).
However, at a low sampling rate where the integral window is unavoidably large, 
the weak formulation's low-pass filtering characteristic leads to noticeable
artificial smoothing in the data (figure \ref{fig:VanDerPolWeakVSFD}
\((a,e)\)). By comparison, the finite difference method remains
reasonably unbiased despite the significant amplification of the noise
variance (figure \ref{fig:VanDerPolWeakVSFD} \((c,g)\)).

Therefore, the decision between weak formulation and finite difference
involves a compromise between attenuating noise and maintaining signal.
The finite difference method does not attenuate any signal from a
spectrum standpoint, but it amplifies noise by the magnitude of the
sampling rate. Hence, it is more suitable for data sampled at a low
sampling rate. Conversely, the weak formulation is a proficient method
for attenuating noise and mitigating noise amplification, but it may
also suppress signals at high frequencies. It is suitable for data that
is sampled at a rate significantly higher than the Nyquist frequency of
the signal. In practice, the sampling rate is often limited by the
capabilities of the measuring hardware, and problems that push the limit
of the equipment are often sampled at a frequency just above the Nyquist
frequency of the relevant dynamics. As a rule of thumb, the weak formulation
usually requires at least 6-12 data points in the integral. Hence,
practically, it can only resolve dynamics 6-12 times slower than the
sampling frequency. By improving model selection performance in the high
noise regime, the Bayesian reformulation presented in this study improves
the overall chances of recovering the correct model even when the
sampling rate is close to the Nyquist frequency of the dynamic, i.e.~in
the low-data regime.

\section{Examples}\label{sec:examples}

This section will showcase the noise robustness and data efficiency of our Bayesian-SINDy algorithm in recovering the desired equations precisely in comparison with the original SINDy algorithm \cite{Brunton2016} that is based on Sequentially Thresholded Least Squares (STLS). 
We will also compare our Bayesian-SINDy algorithm with \texttt{SparseBayes} \cite{Tipping2003} to show the improvements our Bayesian-SINDy has brought to the Bayesian approaches. 
Four examples will be used to compare the three algorithms,
in which we will consider a polynomial library \(\Theta(\mathbf{X})\) up
to third order.
To compare the three algorithms and to ensure the result is not skewed by individual runs, we will vary the sampling frequency and noise variance for the first three synthetic examples, 
and iterate the learning process 10000 times to compare the percentage of runs for which each algorithm has successfully recovered exactly all the terms in the original equation without extra spurious terms,
 i.e. the success rate of recovering the original equations.
In this study, the use of success rate as the metric is justified because we are focused on the most parsimonious and generalisable equations, which are the original equations, precisely.
Furthermore, to avoid miscounting models with the same sparsity pattern but substantially different coefficients, the success rate is only counted when the model coefficient error (MCE = $||\mathbf{w}-\hat{\mathbf{w}}||_2 / ||\hat{\mathbf{w}}||_2$, where $\hat{\mathbf{w}}$ is the truth) is below $25\%$.
As for the fourth example, we will learn the dynamical
equations from real-world measurements. 
Although a ground-truth model is absent, historical uses of the dataset \cite{Hewitt1921,Mahaffy2010,Hirsh2022} have shown that we should be expecting the Lotka-Volterra equations.

We should note that in this section, we do not aim to claim that Bayesian-SINDy is the most effective method for recovering the correct dynamical equations from data. Furthermore, we note that SINDy (STLS) is not the most effective methodology among many SINDy extensions. As we shall discuss later in \S\ref{subsec:limitation}, we expect more computationally-intensive sampling-based methods to be more robust. 
However, by comparing SINDy (STLS) with an equivalently simple and computationally efficient method derived from the Bayesian framework, we will demonstrate that a Bayesian approach is more robust in noisy and low data scenarios, and thereby encourage a Bayesian perspective to other extensions of SINDy.
The comparison with \texttt{SparseBayes}, meanwhile, demonstrates the improvement we have brought to our learning algorithm.
To compare the algorithms properly, we have also taken steps to ensure they are performing the best they can with the appropriate hyperparameters.
Details of the hyperparameters used for each algorithm and in each example are listed in Appendix B.

\subsection{Van der Pol oscillator}\label{subsec:VanDerPol}

\begin{figure}[h]
\hypertarget{fig:VanDerPol}{%
\centering
\includesvg[ width = 1 \columnwidth ]{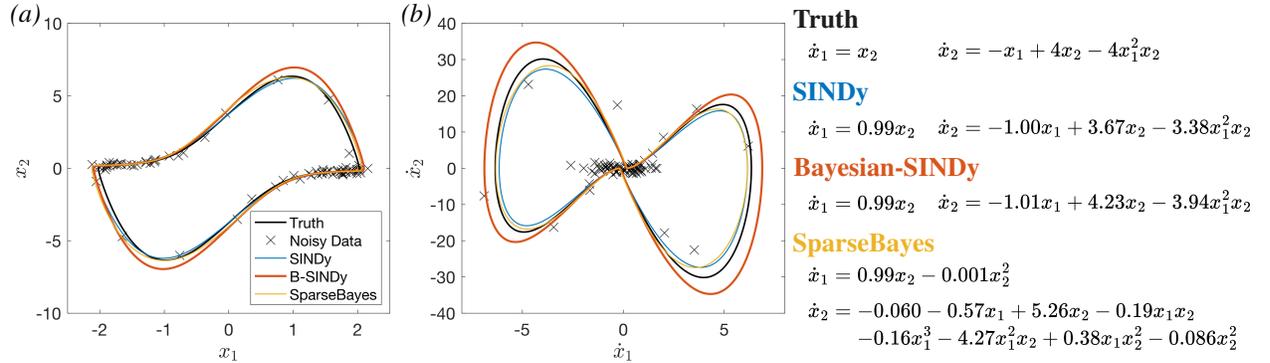}
\caption{The Van der Pol system (\ref{eq:VanDerPol}) with \(\beta=4\) (black line)
and the system estimated by SINDy (STLS, blue), Bayesian-SINDy (red) and \texttt{SparseBayes} (RVM, yellow) given the noisy data of the system (black crosses). The right panel shows the equations learnt, while $(a)$ shows the phase diagram in $x_1-x_2$ space, and $(b)$ shows the equivalent in $\dot{x}_1-\dot{x}_2$ space. Finite difference is used to obtain $\dot{\mathbf{x}}$. 
The noise level is \(\sigma_x=0.1\) and
the data are sampled at \(\Delta t=0.15\) for $t \in [0, 12]$. }\label{fig:VanDerPol}
}
\end{figure}

\begin{figure}[h!]
  \hypertarget{fig:VanDerPolRate}{%
  \centering
  \includesvg[ width = 1 \columnwidth ]{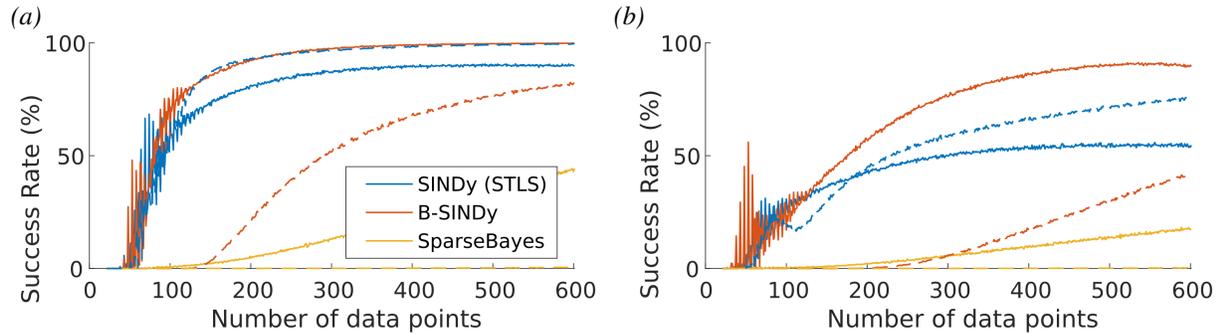}
  \caption{The percentage of runs that successfully identifies the Van der Pol oscillator equations from data using SINDy (STLS, blue), Bayesian-SINDy (red) and \texttt{SparseBayes} (yellow) after 10000 iterations against the number of data points $\tilde{N}=12/\Delta t$. Solid lines show results from finite differences and dashed lines show results from weak formulations. The noise levels are $(a)$ $\sigma_x=0.1$ and $(b)$ $\sigma_x=0.2$. 
  Note that the oscillation at low $\tilde{N}$ is an artefact of stroboscopic effect. }\label{fig:VanDerPolRate}
  }
\end{figure}

In this example, we consider the Van der Pol oscillator system given by
\begin{equation}
\dot{x}_1= x_2, \quad 
\dot{x}_2 = b x_2 (1-x_1^2)-x_1,
\label{eq:VanDerPol}\end{equation} in which we use the values \(b=4\)
and the initial conditions \(\mathbf{x}=[2,0]\). A Gaussian measurement
noise with distribution \(\mathcal{N}(0,0.1^2)\) or \(\mathcal{N}(0,0.2^2)\) is added to the data,
sampled over the interval \(t \in [0, 12]\). Figure
\ref{fig:VanDerPol} shows a typical example of the data provided to the
algorithms and the estimation by SINDy (STLS), Bayesian-SINDy and \texttt{SparseBayes}, 
while figure \ref{fig:VanDerPolRate} shows the success rates of the each algorithm using both an 8th-order (9 points) central difference scheme and a weak formulation scheme with 9 points and $\phi=(t^2-1)^2$.

At both noise levels, Bayesian-SINDy shows robust performances in recovering the Van der Pol equations using finite difference despite the noise amplification (c.f. figure \ref{fig:VanDerPolWeakVSFD}$h$). Although the weak formulation helps SINDy recover the equations when noise is low, its success rate remains lower than Bayesian-SINDy when noise is high. Furthermore, in the high-noise case, the weak formulation causes SINDy's performance to drop below that of finite differences. This is due to the low-pass filtering property, as anticipated by our discussion
in \S \ref{sec:WeakVSFD} and illustrated in figure
\ref{fig:VanDerPolWeakVSFD}. 
Meanwhile, \texttt{SparseBayes} performs consistently worse than the other two algorithms.
The poor performance of \texttt{SparseBayes} is due to the lack of facilities to account for noise variation over time, which we will further explain in \S\ref{sec:CubicOsc}.

One may notice that the success rate is highly oscillatory when $\tilde{N}$ is low. 
This is due to a stroboscopic effect at low sampling rates.
In a stiff system such as the Van der Pol oscillator, the majority of the information about the system is concentrated within the few data points at which the dynamics spike 
(i.e. when $\dot{\mathbf{x}}$ is far from $\mathbf{0}$ in figures \ref{fig:VanDerPolWeakVSFD}$(e-h)$ and \ref{fig:VanDerPol}$(b)$). 
The rest of the data points simply repeat information about the plateau region, where $\dot{\mathbf{x}}$ is close to zero.
With only two spikes within the time interval, successful recoveries of the equations are highly sensitive to whether the spikes are sampled sufficiently in the data. With a continuously changing sampling rate and fixed time interval, the stroboscopic effect implies that the spike might not be sampled enough at some sampling rates, leading to oscillation in the resulting success rate. 
Later in \S\ref{subsec:Active}, we will show how to drive the algorithms to preferentially learn from more informative data points and improve data efficiency.

\subsection{Cubic oscillator}\label{sec:CubicOsc}

In this example, we consider the damped nonlinear oscillator given by
\begin{equation}
\dot{x}_1= a x_1^3+b x_2^3, \qquad 
\dot{x}_2 = c x_1^3 + d x_2^3,
\label{eq:CubOsc}\end{equation} in which we use the values
\(a=-0.1,b=-2,c=2,d=-0.1\), and the initial conditions
\(\mathbf{x}=[1,0]\). Data are sampled over a fixed time interval
\(t \in [0, 5]\) at various rates, and a Gaussian measurement noise with distribution
\(\mathcal{N}(0,0.005)\) or \(\mathcal{N}(0,0.01)\) is added to the data. To recover
the time derivatives, here we use the weak formulation with 7 data
points, with a test function \(\phi=(t^2-1)^4\).

\begin{figure}[h!]
  \hypertarget{fig:CubOsc}{%
  \centering
  \includesvg[ width = 0.99 \columnwidth ]{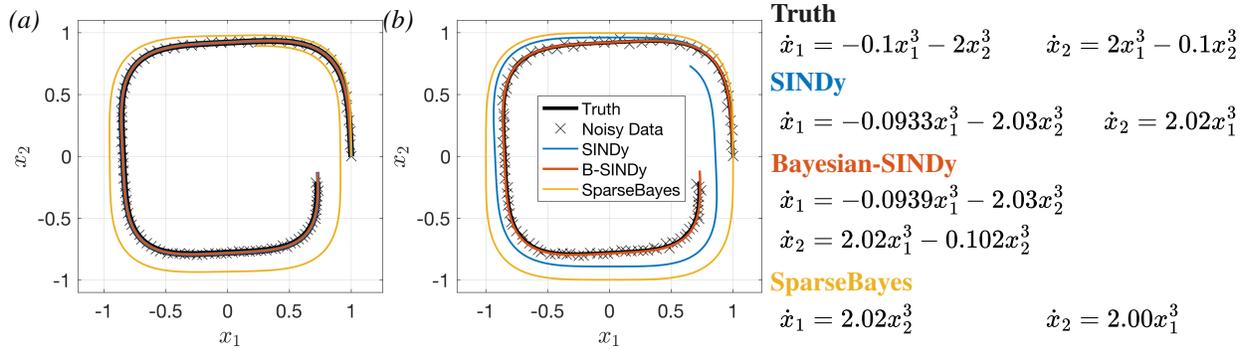}
  \caption{The phase diagram of the cubic nonlinear oscillator system (\ref{eq:CubOsc}, black line) and the system estimated by SINDy (STLS) (blue), Bayesian-SINDy (red) and \texttt{SparseBayes/RVM} (yellow) given the noisy data (crosses) of the system. The noise level are $(a)$ $\sigma_x=0.005$ and $(b)$ $\sigma_x=0.01$ and the data are sampled at $\Delta t=0.05$ for $t \in [0, 5]$. The right panel shows the model equations from each algorithm when $\sigma_x=0.01$. In $(a)$, the prediction by SINDy and B-SINDy is so close to truth their differences are practically indiscernible. }\label{fig:CubOsc}
  }
\end{figure}

\begin{figure}[h!]
  \hypertarget{fig:OscVarRate}{%
  \centering
  \includesvg[ width = 1 \columnwidth ]{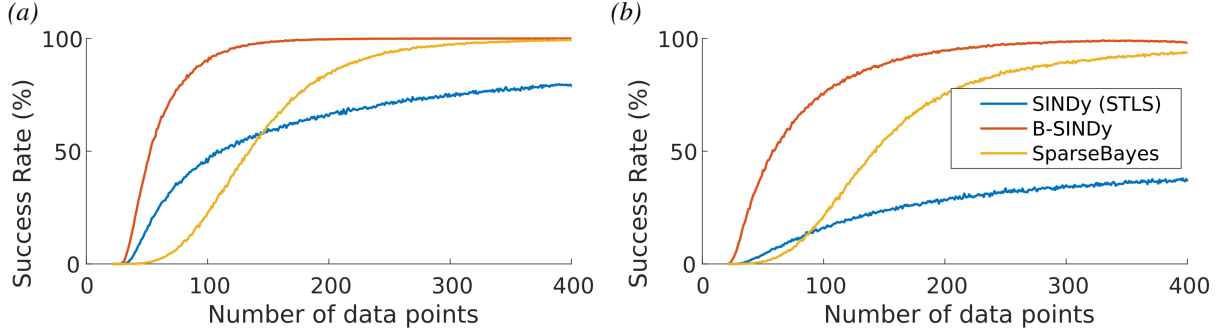}
  \caption{The percentage of runs that successfully identifies the cubic nonlinear oscillator equations from data using SINDy (STLS, blue), Bayesian-SINDy (red) and \texttt{SparseBayes} (yellow) after 10000 iterations against the number of data points $\tilde{N}=5/\Delta t$. The noise levels are $(a)$ $\sigma_x=0.005$ and $(b)$ $\sigma_x=0.01$.}\label{fig:OscVarRate}
  }
\end{figure}

Figure \ref{fig:CubOsc} depicts a typical instance of the data provided
to the algorithms and the resulting learning outcome of Bayesian-SINDy, and
figure \ref{fig:OscVarRate} summarises the success rate of each algorithm in recovering (\ref{eq:CubOsc}). Here,
Bayesian-SINDy has shown better performance in all scenarios examined. \texttt{SparseBayes} is also resilient to noise, but it is less data-efficient than Bayesian-SINDy.
In contrast, SINDy (STLS) exhibits a decline in performance when the noise variance increases.

\begin{figure}
\hypertarget{fig:OscVar}{%
\centering
\includesvg[ width = 1 \columnwidth ]{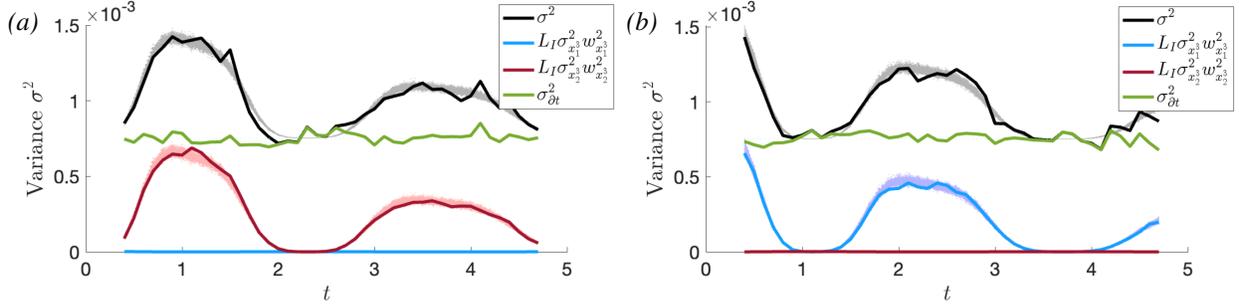}
\caption{Cubic oscillator: breakdown of contributions to the total variance of noise $\boldsymbol\sigma$ in the dynamics of $(a)$ $x_1$ and $(b)$ $x_2$ by each term in (\ref{eq:total_noise}). Thick lines show the variance computed from the statistics
of 1000 iterations, and lighter lines (shaded area) show the bound of the theoretical
calculation using (\ref{eq:total_noise}) from each iteration. Here, $\Delta t =0.1$.}\label{fig:OscVar}
}
\end{figure}

One reason Bayesian-SINDy performs better than \texttt{SparseBayes} is that it properly accounts for the noise variance according to (\ref{eq:total_noise}). 
As shown in figure \ref{fig:OscVar}, the noise variance varies significantly over time due to the contribution from the cubic terms.
However, \texttt{SparseBayes} presumes a constant (inverse) noise variance $\boldsymbol\beta$ over time and hence cannot account for the time-varying contribution from the cubic terms.
In contrast, the noise iteration procedure in Bayesian-SINDy captures the contribution of $\boldsymbol{\sigma}_\Theta^2$ to $\boldsymbol{\sigma}^2$ well, thereby improving Bayesian-SINDy's performance in the low-data limit. 

\subsection{Lorenz system}\label{subsec:Lorenz}

In this example, we consider the Lorenz system given by \begin{equation}
\dot{x}_1= s(x_2-x_1), \qquad 
\dot{x}_2 = r x_1-x_2-x_1 x_3, \qquad 
\dot{x}_3 = x_1 x_2 - b x_3,\\
\label{eq:Lorenz}\end{equation} in which we use the values
\(s=10,b=8/3,r=28\), and the initial conditions
\(\mathbf{x}=[-1,6,15]\). Data are sampled over the time
interval \(t \in [0, 2.5]\), and a Gaussian measurement noise with
distribution \(\mathcal{N}(0,\sigma_x^2)\) is added to the data, with \(\sigma_x\) varying between \(0.025\) to \(0.2\). 
Here, we use a central finite difference with 13 data points to recover the time derivatives. 

\begin{figure}
  \hypertarget{fig:LorenzRate}{%
  \centering
  \includesvg[ width = 0.9 \columnwidth ]{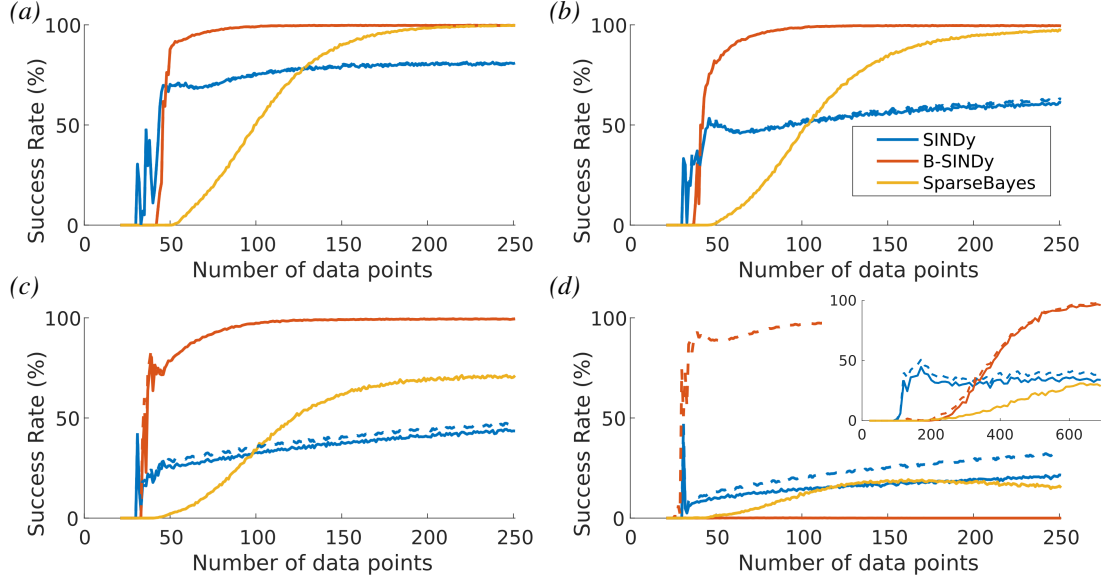}
  \caption{The percentage of runs that successfully identifies the Lorenz system equations from data sampled between $t \in [0,2.5]$ using SINDy (STLS, blue), Bayesian-SINDy (red) and \texttt{SparseBayes} (yellow) after 10000 iterations against the number of data points $\tilde{N}=2.5/\Delta t$. 
  The noise levels are $(a)$ $\sigma_x=0.025$, $(b)$ $\sigma_x=0.05$, $(c)$ $\sigma_x=0.1$ and $(d)$ $\sigma_x=0.2$. The dashed line shows the success rate of recovering the model without the \(x_2\) term in the equation for \(\dot{x}_2\). The inset in $(d)$ shows the success rate when the time interval is extended to $t \in [0,10]$.}\label{fig:LorenzRate} 
  }
\end{figure}

Figure  \ref{fig:LorenzRate} shows the success rates of the three algorithms.
When noise is low, SINDy shows moderate success in recovering the dynamical equations, but performance degrades quickly as noise increases.
Meanwhile, both Bayesian-SINDy and \texttt{SparseBayes} show more robustness in learning the correct model as noise increases, and Bayesian-SINDy maintains better performance at the low-data limit.

\begin{figure}[h!]
\hypertarget{fig:LorenzX2}{%
\centering
\includesvg[ width = 0.75 \columnwidth ]{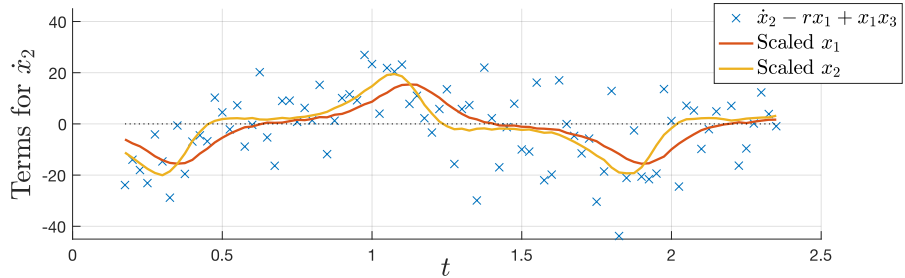}
\caption{Learning the dynamics of $\dot{x}_2-rx_1+x_1x_3$ in the Lorenz system: the crosses show the values of $\dot{x}_2-r x_1 +x_1x_3$ from the data, which should match $x_2$ (yellow line). However, $x_1$ (red line) is highly collinear with $x_2$, and it is not obvious which term should fit the data better. Both $x_1$ and $x_2$ are scaled to best match the data $\dot{x}_2-r x_1 +x_1x_3$. Here, $\Delta t=0.025$ ($\tilde{N}=100$) and $\sigma_x=0.2$.}\label{fig:LorenzX2}
}
\end{figure}

However, in figure \ref{fig:LorenzRate}$(d)$, in which the noise is $\sigma_x=0.2$, none of the three algorithms can recover the model exactly. 
Both SINDy (blue) and \texttt{SparseBayes} (yellow) can only recover the full model sporadically and showed little consistency in the model they produced. 
In contrast, Bayesian-SINDy consistently recovers a model with only the \(x_2\) term in equation for \(\dot{x}_2\) missing (as shown by the dashed red line).
The missing \(x_2\) term in \(\dot{x}_2\) is
particularly difficult to learn because it is highly collinear with the existing $x_1$ term in the model. 
To distinguish the two would require more data in regions in which the two differ significantly, but, as shown in figure \ref{fig:LorenzX2}, the given signal is highly obscured by noise and does not sample those regions sufficiently in the short time interval \(t \in [0, 2.5]\). 
In other words, there is simply not enough information in the small time window to infer the \(x_2\) term in the equation for \(\dot{x}_2\).
From the evidence-maximisation perspective, this difficulty is reflected in figure \ref{fig:Jalpha}$(b)$, where \(\mathcal{J}(\alpha_m)>0\) only for a small range of \(\alpha_m\).  
In practice, when the data is too low or too noisy, information content in the data may not be enough to infer the full dynamics.
In such cases, 
one inevitably has to increase the amount of data available by considering a long time interval, say, \(t \in [0, 10]\). 
As demonstrated in the inset of figure \ref{fig:LorenzRate}$(d)$, Bayesian-SINDy can eventually recover the full dynamics most of the time, while SINDy remains less robust in this high noise regime.
This lack of information, as depicted here in figure \ref{fig:LorenzX2}, also hints at the possibility of further improving data efficiency if we are allowed to sample data in the most informative state.
We will discuss this information-theoretical approach further in \S\ref{subsec:Active}.

\hypertarget{real-life-data-lotka-volterra-equation-from-lynx-hare-population-dynamics}{%
\subsection{Lotka-Volterra equation from Lynx-Hare population data}\label{real-life-data-lotka-volterra-equation-from-lynx-hare-population-dynamics}}
 
\begin{figure}[h!]
  \hypertarget{fig:LynxHare}{%
  \centering
  \includegraphics[ width = 1 \columnwidth ]{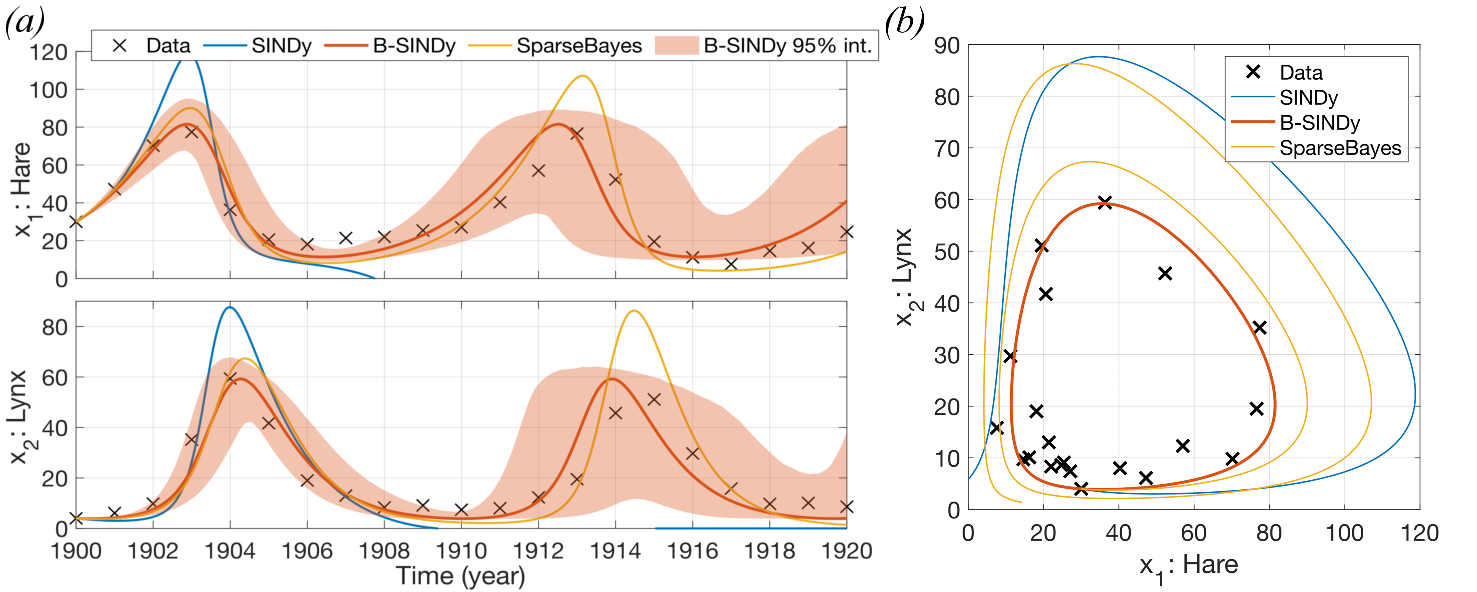}
  \caption{SINDy (STLS, blue), Bayesian-SINDy (red) and \texttt{SparseBayes} (yellow) applied to the lynx-hare population data based on the pelts trade by the Hudson Bay Company between 1900 and 1920 \cite{Hewitt1921,Mahaffy2010}.\\ $(a)$ Hare (upper) and Lynx (lower) populations over time. Solid lines show the model reconstruction from the parameters estimated by each algorithm, and the shaded area depict the $95\%$ confidence interval given by the ensemble reconstruction using the parameter uncertainty estimated by Bayesian-SINDy. $(b)$ Phase diagram of the Lynx and Hare populations.  \label{fig:LynxHare}}
  }
  \end{figure}

In this example, we apply the three algorithms to model the population
dynamics of Lynx and Hare in Canada, using pelt trade data collected by
the Hudson Bay Company between 1900 and 1920
\cite{Hewitt1921,Mahaffy2010}. Assuming the pelts of these two
species traded by the Company are proportional to the true population, it
is expected that the pelt trade data will approximately adhere
to the Lotka-Volterra model for predator-prey dynamics. In fact, this
dataset has been used as empirical evidence to support the
Lotka-Volterra equations, and has become increasingly popular as a
real-life benchmarking dataset among statisticians and data scientists
\cite{Carpenter2018,Fasel2022,Hirsh2022}. The primary challenge posed by
this dataset is its limited number of data points - specifically, 21
annual data points for each population, resulting in a total of 42 data
points. Due to the low sampling rate, the finite difference method is
more appropriate, but it will, as one expects, also amplify the noise in
the derivative, which we will then mitigate by relying on the noise
robustness of the Bayesian framework. The weak formulation was also
tried, but the low-pass filtering effect filtered vital signals in
the data, leading to subpar learning performance across all methods.

The data is shown in figure \ref{fig:LynxHare}, where \(x_1\)
corresponds to Hare populations, and \(x_2\) corresponds to Lynx
populations. Given the data is not a direct measurement of the
population, it is expected to be highly uncertain. However, the level of
uncertainty is not quantified. The lack of uncertainty quantification is
not a major challenge for SINDy or \texttt{SparseBayes}, as SINDy does
not consider this factor, while \texttt{SparseBayes} can optimise the
noise variance through evidence maximisation. Our Bayesian-SINDy algorithm,
on the other hand, requires a value for \(\sigma_x\).

\begin{table}[h]
  \centering
  \caption{Coefficients and their uncertainty (one standard deviation) estimated from the lynx-hare data using SINDy (STLS), Bayesian-SINDy and \texttt{SparseBayes}. 
  The library consists of all combinations of $x_1$ and $x_2$ up to third order, but the table only shows terms that have non-zero coefficients from one of the algorithms. The non-zero terms in the Lotka-Volterra model are highlighted in yellow.
  }
  \label{tbl:LynxHare}
  \begin{tabular}{|r|cccc |}
 \hline
                 & $\dot{x}_1:1$ & \cellcolor{yellow}$\dot{x}_1:x_1$ & $\dot{x}_1:x_2$                         & \cellcolor{yellow}$\dot{x}_1:x_1 x_2$       \\
  \hline
  SINDy (STLS)   & -9.92       & \cellcolor{yellow}0.82   & 0.31  & \cellcolor{yellow}-0.036                                    \\
  Bayesian-SINDy & 0  & \cellcolor{yellow}0.53 $\pm$ 0.0043          & 0                                       & \cellcolor{yellow}-0.026 $\pm$ 6.1$\times 10^{-6}$                       \\
  \texttt{SparseBayes}  & 0             & \cellcolor{yellow}0.56 $\pm$ 0.0045        & 0                                       & \cellcolor{yellow}-0.027 $\pm$ 5.3$\times 10^{-6}$                              \\
  \hline \hline
                 & $\dot{x}_2:1$                          & \cellcolor{yellow}$\dot{x}_2:x_2$ & \cellcolor{yellow}$\dot{x}_2:x_1 x_2$ & $\dot{x}_2:x_2^3$       \\
  \hline
  SINDy (STLS)   & -1.26                                              & \cellcolor{yellow}-0.92         & \cellcolor{yellow} 0.027             & 0                        \\
  Bayesian-SINDy & 0                                                  & \cellcolor{yellow}-0.98$\pm$ 0.014          & \cellcolor{yellow} 0.028 $\pm$ 1.3$\times 10^{-5}$              & 0                       \\
  \texttt{SparseBayes}  & 0                                            & \cellcolor{yellow}-1.00 $\pm$ 0.018         & \cellcolor{yellow} 0.026 $\pm$ 8.9$\times 10^{-6}$             & $3.6 \times 10^{-5} \pm 1.7 \times 10^{-9}$ \\
  \hline 
  \end{tabular}
\end{table}

To address the difficulty, we ran a hyperparameter sweep on
\(\sigma_x\), assuming \(\alpha=10^{-2}\), and chose the value of
\(\sigma_x \approx 2.7\) by evidence maximisation. Although this
technique is similar to the evidence maximisation process in
\texttt{SparseBayes}, a notable distinction is that our procedure does
not assume a uniform \(\boldsymbol\sigma\) for all data
points. Instead, we calculate it using equation (\ref{eq:total_noise})
and algorithm 1. The parameters estimated are displayed in table \ref{tbl:LynxHare}, where
the Lotka-Volterra model is successfully recovered by Bayesian-SINDy,
while the other two struggle to learn a consistent model. Uncertainties in the estimated parameters, also shown in table \ref{tbl:LynxHare}, are also easily recovered from the posterior distribution. Figure \ref{fig:LynxHare} also shows how the uncertainty in the estimated parameter propagates through the dynamics, assuming the same initial condition. Notably, our
method was able to estimate the parameter uncertainties without expensive sampling, unlike previous techniques
\cite{Hirsh2022}.

\section{Discussion}\label{sec:discussion}

\subsection{\texorpdfstring{Scaling of likelihood and Occam factor with $N$ and $\boldsymbol\beta$}{Why is the Bayesian approach better at the low data limit? Scaling of likelihood and Occam factor with N and beta}}\label{subsec:scaling}

To better understand why the Bayesian-SINDy, and Bayesian approaches in
general, are more robust than frequentist approaches such as STLS at
selecting the correct model in the noisy and scarce data limit, we
compare the two approaches under a unifying Bayesian framework. From the
Bayesian viewpoint, frequentist approaches seek to maximise the
log-likelihood, which, under the premise of Gaussian likelihood, can be
expressed as 
\begin{equation}
\mathcal{J}(\mathbf{y}|\mathbf{D},\mathbf{w},\boldsymbol\beta) = -\frac{N}{2} \ln{2\pi} - \frac{1}{2} ||\boldsymbol\beta||_1 - \frac{1}{2} \boldsymbol\beta \cdot (\mathbf{y}-\mathbf{D}\mathbf{w})^2. \label{eq:log-likelihood}
\end{equation} 
Meanwhile, our Bayesian approach aims to maximise the log-evidence
\begin{subequations}
\begin{equation}
\mathcal{J}(\mathbf{y}|\mathbf{D},\boldsymbol\beta,\boldsymbol\alpha) =
\mathcal{J}(\mathbf{y}|\mathbf{D},\mathbf{w},\boldsymbol\beta) |_{\mathbf{w}=\boldsymbol{\mu}}+
\mathcal{J}(\mathbf{w}|\boldsymbol\alpha)|_{\mathbf{w}=\boldsymbol{\mu}}+\frac{1}{2} \ln{(\det(\boldsymbol\Sigma))},
\label{eq:log-evidence-breakdown}
\end{equation} 
which is composed of three components: the determinant of the covariance matrix 
\begin{equation}
\boldsymbol\Sigma =(\mathbf{A} + \mathbf{D}^T\mathbf{B}\mathbf{D})^{-1} \in \mathbb{R}^{M\times M},
\end{equation}
\end{subequations}
the log-likelihood
\(\mathcal{J}(\mathbf{y}|\mathbf{D},\mathbf{w},\boldsymbol\beta)\), and
the log-prior
\(\mathcal{J}(\mathbf{w}|\boldsymbol\alpha)\) at
the \emph{maximum a posteriori} (MAP) point. The MAP point refers to the parameter values \(\mathbf{w}=\boldsymbol{\mu}\) that maximise the posterior.
\begin{figure}[h!]
  \hypertarget{fig:LorenzRainfall}{%
  \centering
  \includesvg[ width = 0.6 \columnwidth ]{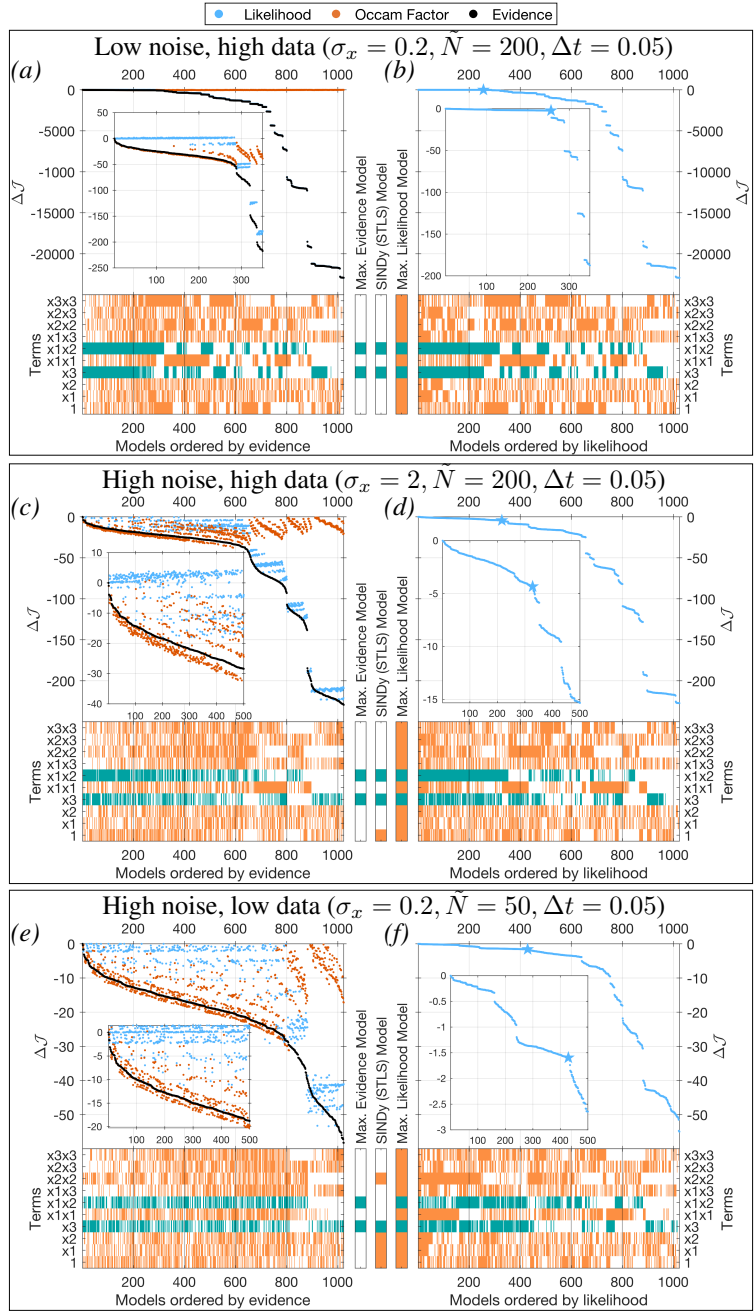}
  \caption{Difference in log-evidence, log-likelihood and the Occam factor (i.e. log-evidence minus log-likelihood) compared to the highest-ranked model of all possible models for $\dot{x}_3$ in the Lorenz system. Models are ranked by $(a,c,e)$ evidence and $(b,d,f)$ likelihood. Insets show a zoomed plot of the first few hundred models, and the barcode plots below show the non-zero terms in each model, marked as green (correct) or orange (incorrect). The correct model is always ranked highest by evidence in $(a,c,e)$, and indicated with a star in $(b,d,f)$. Meanwhile, the model selected by highest likelihood contain all the terms, and the models selected by SINDy are shown in the middle. Here, $(a-b)$ $\sigma_x=0.2$ and $\tilde{N}=200$, $(c-d)$ $\sigma_x=1$ and $\tilde{N}=200$, and $(e-f)$  $\sigma_x=1$ and $\tilde{N}=50$. }\label{fig:LorenzRainfall}
  }
\end{figure}
It should be noted that in equation (\ref{eq:log-likelihood}), the log-likelihood scales linearly with the magnitude of the inverse noise variance \(\boldsymbol \beta\) and the number of data points \(N\). 
On the other hand, the remaining components in the log-evidence, also known as the
Occam factor, serve as a penalty for models that have an excessive
number of parameters. The Occam factor scales with the number of terms
\(M\) in the model and only has a partial and weak \(\ln{N}\) scaling
with \(N\).

Figure \ref{fig:LorenzRainfall} demonstrates how the scaling of the  log-likelihood and the Occam factor plays out in the model selection process.
The left column $(a,c,e)$ of figure \ref{fig:LorenzRainfall} shows the contribution of the log-likelihood (blue)
and the Occam factor (red) to the log-evidence across all 1024 possible
models for $\dot{x}_3$ in the Lorenz system when a smaller, second-order polynomial library containing ten terms is considered. 
For each model, the barcode plot below show the library terms that are non-zero, with the non-zero terms $x_3$ and $x_1 x_2$ in the true model coloured green.
The models are sorted by log-evidence, demonstrating a Bayesian perspective on model selection, in which the model with maximum evidence is selected as the most parsimonious model for the data. The maximum evidence model (Bayesian-SINDy) identifies the correct model in all three scenarios.  

For the same dataset, the right column $(b,d,f)$ of figure \ref{fig:LorenzRainfall} sorts the model based on log-likelihood calculated from the maximum likelihood estimate (MLE), i.e. $\mathbf{w}=\boldsymbol{\mu}_{\text{MLE}}$, which can be obtained by ordinary least squares. (Note that when $N$ and $\beta$ is large, the MLE and MAP estimates of $\boldsymbol{\mu}$ converge.)
In contrast to the Bayesian perspective on the left, the right column reflects a frequentist perspective in which the model with the best fit is selected as the best model for the data. 
As one would expect, the MLE-based model selection always selects the model with all ten terms (i.e. over-fit). 
To induce sparsity, the frequentist approach STLS introduces an empirical thresholding process. The idea is to remove small coefficient terms that don't contribute much to the likelihood, thereby achieving a compromise between maximising likelihood and sparisty.

When the data is abundant and the
noise is minimal (i.e.~\(N ||\boldsymbol\beta|| \gg M\)), models that
include the correct terms have significantly higher likelihood and are
easily distinguishable from the other models when sorted by likelihood.
This is demonstrated by the continuous green band in the barcode plot in \ref{fig:LorenzRainfall}$(b)$.
Phenomenologically, this signifies that all models containing the
appropriate terms exhibit comparable levels of likelihood. STLS
leverages this clear separation in likelihood and only eliminates irrelevant terms that do not affect the likelihood much in its thresholding process,
thereby effectively constraining its search for
the most sparse model within this region of large likelihood.

However, as the amount of data is lowered (i.e.~smaller \(N\)), or when
the noise increases (i.e.~smaller \(||\boldsymbol\beta||\)), the distinct
separation of the high likelihood region from the other possible models
becomes less evident. In these scenarios, models lacking the relevant
terms are more likely to fit the data incidentally, as reflected by the
discontinuous green band in the barcode plot of figure \ref{fig:LorenzRainfall}$(d,f)$. STLS may,
thus, converge on an incorrect model. Alternatively, STLS may not
converge to the most sparse model because there is a non-negligible chance that
irrelevant terms will also have coefficients larger than the threshold
when \(N\) or \(\beta\) decreases.

Conversely, figure \ref{fig:LorenzRainfall}$(a,c,e)$
demonstrates that the Occam factor assumes greater significance in model
selection when noise is large and data is scarce. The Occam factor
serves two crucial functions in model selection. First, it penalises
models with too many terms because it scales with \(M\). Second, it
quantifies the variance of the estimated parameters, and discards terms that
have small variances in their coefficients because this indicates that the
corresponding model is not robust (or is too sensitive to small changes in
the parameter values). In performing the above two functions, the
rigorously derived Occam factor excludes terms that incidentally fit the
data, ensuring the robustness of the estimated model. 

\subsection{Limitations and applications of Bayesian-SINDy\label{subsec:limitation}}
One of the key contributions of this study is the development of a computationally efficient estimation of the noise variance, 
which is critical in calculating the evidence Bayesian-SINDy optimizes for during the model selection process.
Mutliple approximations were made as a trade-off between computational efficiency and accuracy, 
which may limit the applicability of Bayesian-SINDy despite its greater resilience than the original STLS-based SINDy in accurately identifying the proper dynamical equations from noisy and scarce datasets.
Therefore, here we provide a review of the approximations made, their limitations, and potential remedies.

\paragraph*{Gaussian noise} The noise model for $\boldsymbol\epsilon_x$ is assumed to be Gaussian, which is a common assumption in many statistical methods and is chosen in this study mainly for its tractability. Real-world data may not, however, follow a Gaussian distribution. Nonetheless, the framework for the noise estimation laid out in \S\ref{sec:SINDy} is built upon a Taylor expansion of moments, which is not restrictive to Gaussian noise. Therefore, this framework can be extended to other noise distributions by modifying the noise estimation procedure in \S\ref{sec:SINDy} to account for the different moments of the noise distribution.

\paragraph*{Noise correlation in time} One major assumption in the noise model is that the noise is uncorrelated in time. This assumption was made mostly to speed up computation, but is not neccessary. In fact, one can easily relax the assumption by replacing the inverse variance $\boldsymbol\beta$ with the full inverse covariance matrix $\mathbf{B}$. The downside of this approach is that the computational cost of the regression procedure would increase from $\mathcal{O}(N)$ to $\mathcal{O}(N^3)$, which may be prohibitive for large datasets.

\paragraph*{Noise correlation between terms} Another assumption used to speed up computation was that the noise is uncorrelated between terms. If the noise correlation between terms is strong, the noise estimation may not be accurate. Unfortunately, there is no remedy to this issue without incurring significant computational cost, as there are combinatorially many correlations between each term to be considered while the algorithm selects the best model. However, one can always check the validity of the noise model by comparing the estimated noise variance with MCMC \emph{a posteriori}.

\paragraph*{Estimating the noise distribution from the nonlinear functions} As mentioned in \S\ref{sec:SINDy}, as the noise propagates through the nonlinear functions in $\boldsymbol\Theta(\mathbf{X})$, the resulting noise distribution is non-Gaussian. While we use a Taylor expansion to approximate the resulting variance, this estimation is only valid when the noise is relatively small such that higher order terms can be neglected.
\begin{figure}[t]
  \hypertarget{fig:NonlinearOscDist}{%
  \centering
  \includesvg[ width = 0.7 \columnwidth ]{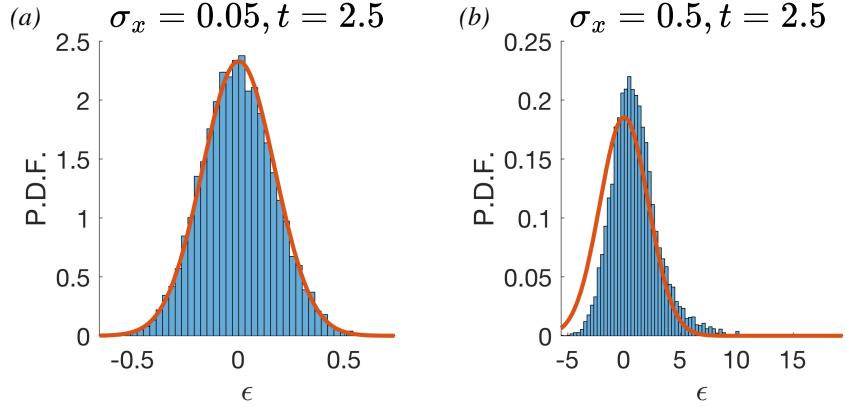}
  \caption{Distribution of the total noise $\boldsymbol\epsilon$ for $x_2$ at $t=2.5$ in the cubic oscillator example with $\Delta t=0.1$. Here, the distribution is created from 10000 samples with noise level $(a)$ $\sigma_x=0.05$ and $(b)$ $\sigma_x=0.5$. The histogram shows the distribution from data, while the red line shows the Gaussian estimation using variance $\boldsymbol\sigma^2$ from (\ref{eq:total_noise}). Notice that the distribution is skewed and deviate from the Gaussian distribution when the noise is too big.}\label{fig:NonlinearOscDist}
  }
\end{figure}
For example, in the nonlinear oscillator case, if one further increases the noise, none of the three algorithms will be able to recover the correct equations. 
This is because, in the presence of highly nonlinear cubic terms, 
the noise distribution can become highly skewed, as shown in \ref{fig:NonlinearOscDist}, breaking the variance and mean approximation from the Taylor expansion of moments.
To capture the skewed distribution, one must extend the algorithm to consider higher order moments, incurring significant computational cost.
Given the restriction in noise level under this approximation, a user may want to gauge the validity of this approximation. 
One way to probe its validity \emph{a priori} is to sample around the data with the given noise level and pass the sample through the nonlinear functions. 
By comparing the skewness and conformity of the resulting distribution with what would have been expected from the approximation (i.e., a Gaussian distribution), 
one can determine if the proposed algorithm provides suitable approximation of the noise distribution.
If the application is not time-critical, it is also a good practice to check the generated model against MCMC sampling \emph{a posteriori} to ensure the noise model's validity.

\paragraph*{Gaussian approximation of the likelihood}
In \S\ref{sec:SINDy}, after propagating the noise through the transformation in (\ref{eq:dyn_dis_noise}) by tracing the transformation on the noise variance $\boldsymbol\sigma^2$, we arrive at the total noise $\boldsymbol\epsilon$ in (\ref{eq:tDw}) with the knowledge of its variance. 
We then approximate the likelihood to be Gaussian with variance $\boldsymbol\sigma^2$. 
This final Guassian approximation is what enables the proposed regression-based algorithm and \texttt{SparseBayes} to be fast. 
This approximation can be justified by the fact that, in the context of performing type-II maximum likelihood optimisation, a Laplace's approximation to the evidence is usually good enough to rank models in the correct order.
The Laplace's approximation only requires the knowledge of the \emph{Maximum a Posteriori} (MAP) probability and the Hessian around the MAP point, which can be approximated as the variance of the posterior,
and is fairly accurate as long as the posterior is unimodal and does not have a heavy tail.

\paragraph*{Applications suitable for Bayesian-SINDy}
Although the computational efficiency of Bayesian SINDy requires many approximations, the algorithm still selects  the correct models robustly and accurately in many cases, as demonstrated empirically in \S\ref{sec:examples}.
However, when the noise distributions are no longer tractable or well approximated by the proposed method, e.g. when the signal-to-noise ratio is too low or the noise is too skewed,
it is recommended to resort to sampling-based techniques to better capture the accurate noise distribution. 
Recent work by Hirsh \emph{et al.} \cite{Hirsh2022} has demonstrated the possibility of further improving noise robustness and
data efficiency by accurately capturing distributions through sampling techniques such as MCMC.
However, this improvement comes at the expense of costly sampling. 
For example, it took hours of computation to recover the
Lotka-Volterra equation from the Hudson Bay Lynx-Hare population dataset
using the MCMC-based method in \cite{Hirsh2022}, compared to the
sub-second computation by Bayesian-SINDy.
Even if a further \emph{a posteriori} check is performed using MCMC, the overall computation time would still be significantly lower than Hirsh's method \cite{Hirsh2022}.
Consequently, the choice of which approach to adopt depends on the specific applications and circumstances. 
If computing time and cost are not significant constraints, sampling-based methods such as those suggested
by \cite{Martina-Perez2021} or \cite{Hirsh2022} should
give more accurate models than Bayesian-SINDy. 
Conversely, if time or computational cost is a major concern, Bayesian-SINDy is a valuable tool for conducting rapid preliminary
investigations. 
In situations that require rapid learning, such as system
identification in control, using a sampling-based method would not be practical. 
Instead, Bayesian-SINDy could be a cost-effective approach to extracting symbolic models from noisy and limited data in a fast and reliable manner.

\subsection{Application of Bayesian-SINDy in active learning \label{subsec:Active}}
One way to further improve data efficiency in the Bayesian framework is to optimise experimental design. 
Lindley \cite{Lindley1956} established the approach to optimal design by choosing, from all possible data points (labelled `experiments' in \cite{Lindley1956}), the one that maximises the information content. 
This information content is quantified by a utility function akin to Shannon's information entropy, which is naturally defined by the probabilistic description in the Bayesian framework. 
However, evaluating the information entropy usually involves integration in high-dimensional spaces, which can be expensive in a sampling-based technique. 
For example, in a parameter inference problem, the information entropy is defined by the parameter posterior. 
Evaluating this information entropy would require integration over the entire parameter space, the cost of which scales exponentially with the number of parameters.

This computational complexity is further exacerbated if an information-based approach analogous to Lindley's is applied to the model selection problem. The information entropy $H$ of interest here is 
\begin{equation}
H = - \int p(\mathbf{y}|\mathbf{D},\boldsymbol\beta,\boldsymbol\alpha) \ln{(p(\mathbf{y}|\mathbf{D},\boldsymbol\beta,\boldsymbol\alpha))} d\mathbf{y}, \label{eq:entropy}
\end{equation}
which is based on the evidence $p(\mathbf{y}|\mathbf{D},\boldsymbol\beta,\boldsymbol\alpha)$. 
In a sampling-based technique, the evaluation of $p(\mathbf{y}|\mathbf{D},\boldsymbol\beta,\boldsymbol\alpha)$ already requires the integration of the posterior in the parameter space. 
The evaluation of $H$, then, would have required further integration in the $\mathbf{y}$, which would be prohibitively expensive if a sampling technique were used. 
Fortunately, the expensive evaluation can be avoided if we accept the Gaussian approximation used throughout this study. 
Then, the analytical tractability of the Gaussian distribution allows for $H$ to be rewritten explicitly as 
\begin{equation}
  H = \frac{N}{2}(1+\ln{(2\pi)}) + \frac{1}{2}\ln{|\mathbf{C}|}, \label{eq:entropyC}
\end{equation}
after substituting (\ref{eq:log-evidence}-\ref{eq:C_def}) into (\ref{eq:entropy}). Here, $\mathbf{C}$ is the covariance matrix of the evidence.

However, in contrast to Lindley's approach, which seeks to maximise information content (analogous to the negative of $H$), here we seek to maximise the information entropy $H$ over all the possible data we can sample. This might seem counter-intuitive, but unlike the parameter inference problem, in which one seeks to minimise uncertainty in the parameters, in our model selection problem, one seeks to minimise certainty of the current model so as to encourage the exploration for a better model. This exploration maximisation through entropy is analogous to entropy maximisation in reinforcement learning. 

Optimising data efficiency has, therefore, become an exercise of sampling data that maximises the covariance $\mathbf{C}$. 
To illustrate the potential improvement this may bring, here we consider again the Van der Pol oscillator and the Lorenz system as examples. 
Instead of using the entire dataset, here, each new data point is selected by maximising gain in $H$ given the estimated model from the existing dataset. 
Then the newly selected point is assimilated, and the model is updated with the new dataset. This process is repeated, and the number of data points required to recover the full model is recorded.

\begin{figure}[h!]
  \hypertarget{fig:VanDerPolEntropy}{%
  \centering
  \includesvg[ width = 0.95 \columnwidth ]{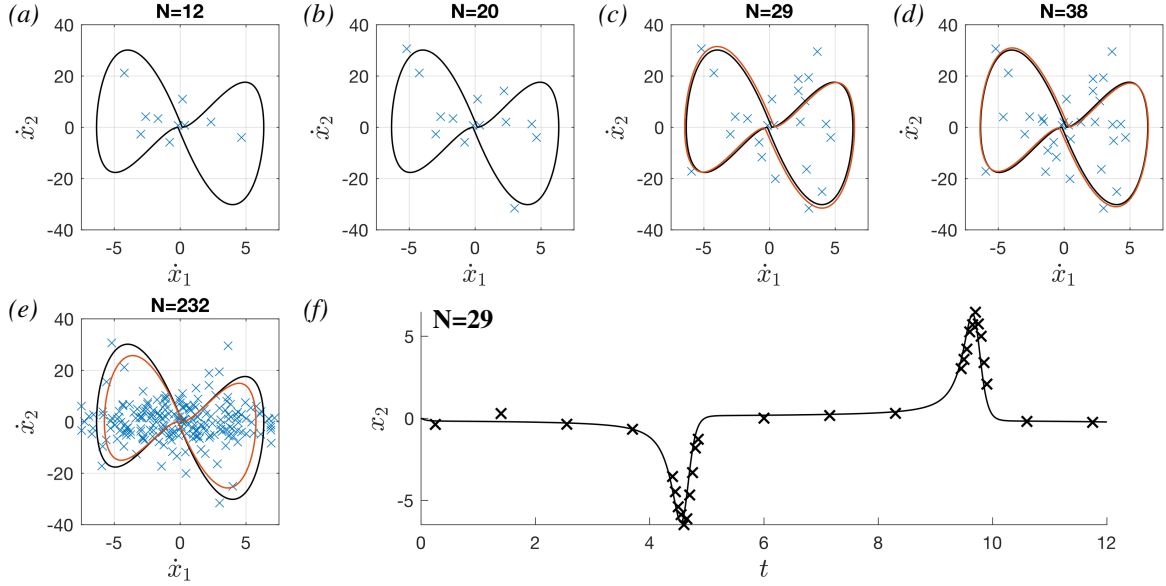}
  \caption{Recovery of the Van der Pol model by sequential data assimilation, where each new data point is selected by maximising information gain. Blue crosses in $(a-e)$ show the data sampled in the phase diagram of $\dot{\mathbf{x}}$, and black lines show the original dynamic without noise. At $N=29$ (panel $(d)$), the full model is recovered, and the red dashed line shows the prediction by the recovered model. Panel $(e)$ shows the full dataset with $N=232$ points. Panel $(f)$ shows the data sampled on $x_2-t$ space when $N=29$, which are concentrated at the region when the dynamic spikes.  }\label{fig:VanDerPolEntropy} 
  }
\end{figure}

In the Van der Pol example, we have briefly mentioned that the majority of the system's information is concentrated during the short time when the dynamics $\dot{\mathbf{x}}$ is far from the origin (see \S\ref{subsec:VanDerPol}). Here, figure \ref{fig:VanDerPolEntropy} shows how effective the information gain $H$ is in guiding the learning algorithm to concentrate sampling in the more informative regions, thereby recovering the full system with less data. 
The full system is recovered with only $N=29$ data points, an order of magnitude lower than the full dataset.  Notice that most data are sampled around the spikes in figure \ref{fig:VanDerPolEntropy}$(f)$, showing that most information about the system is indeed concentrated in those regions. 

\begin{figure}[h!]
  \hypertarget{fig:LorenzEntropy}{%
  \centering
  \includesvg[ width = 0.85 \columnwidth ]{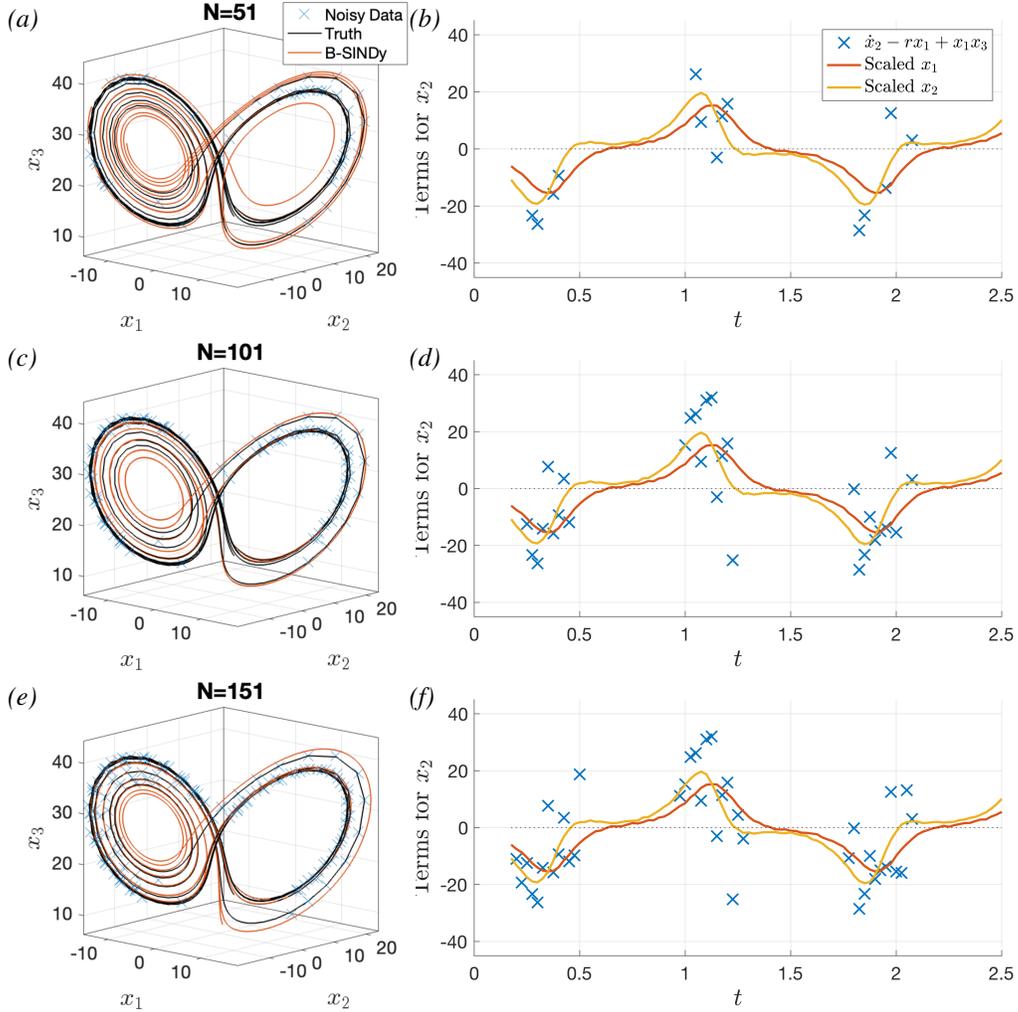}
  \caption{Recovery of the Lorenz model by sequential data assimilation, for which each new data point is selected by maximising information gain. In $(a,c,e)$, blue crosses show the data sampled, and black lines and red dashed lines show the original dynamic without noise and the model prediction, respectively. Similar to figure \ref{fig:LorenzX2}, panels $(b,d,f)$ show the residue signal from the dynamics of $\dot{x}_2$, which should be fitted by $x_2$. Here, blue crosses show the data sampled, which is concentrated at the time where $x_2$ is the most distinguishable from $x_1$.   
  At $N=51$ and $N=101$ (panels $(a-d)$), the Lorenz model is partially recovered, with the exception of the $x_2$ term in $\dot{x}_2$.
  At $N=151$ (panels $(e-f)$), the full model is recovered. }\label{fig:LorenzEntropy} 
  }
\end{figure}

As for the Lorenz system, \S\ref{subsec:Lorenz} has shown that the $x_2$ term in $\dot{x}_2$ is particularly difficult to learn due to high noise that obscurs the dynamics and a lack of information to distinguish $x_2$ from $x_1$. Here, figure \ref{fig:LorenzEntropy} shows how $H$ preferentially samples regions with more information to distinguish between the two terms, thereby significantly reducing the overall data required to recover the full model.

In real-life applications such as system identification for control, one can imagine using $H$ as part of an objective function for optimal control. By balancing the objective of learning more about the dynamics governing the system (i.e. exploration) with the objective of achieving certain target state(s) (i.e. exploitation), one can develop the current framework into a fully automated active learner that can learn about the system and rapidly drive it to a certain state. 

We also note that, as well as our current definition of $H$, there exists a whole family of strategies in optimal experimental design theory to optimise model selection in the most data-efficient manner, yet its application within the SINDy framework remains largely unexplored due to a lack of probabilistic interpretation of the learning result of SINDy. We hope that by grounding SINDy in the Bayesian framework, we can encourage further exploration for more optimal strategies to learn about the system in the most data-efficient manner.

\section{Conclusion and future work}
In this study, we have presented a unifying Bayesian framework to compare Bayesian-SINDy with the original STLS-based SINDy. By grounding STLS in the Bayesian framework, we have revealed how STLS can be a robust model selection method in the high-data low-noise limit due to the dominant contribution of the likelihood to the evidence. However, in the low-data high-noise limit, the Occam factor becomes more important, and Bayesian-SINDy's ability to penalise overfitting becomes crucial. 
The Bayesian framework also provides a more rigorous justification for selecting more parsimonious models than the STLS-based SINDy. 
Instead of selecting models by heuristically selecting hyperparameters at the Pareto front, the Bayesian framework formally defines the Occam factor, resulting in a more principled approach to model selection. 
This results in the Bayesian-SINDy algorithm, which provides a more concrete theoretical foundation for current SINDy-based approaches and a more robust framework for further extensions. 
Along with other recent publications highlighting the Bayesian perspective on SINDy, such as the demonstration of the convergence of Ensemble-SINDy to Bayesian statistics found in \cite{Gao2023}, we conclude that many current SINDy extensions could benefit from the Bayesian framework to improve robustness in noisy and scarce datasets.

To demonstrate how a rigorously defined Occam factor helps model selection in the low data regime, we substituted STLS with sparse Bayesian learning to create the Bayesian-SINDy algorithm. Notably, in all examples considered, the Bayesian-SINDy algorithm was able to recover the intended equations most of the time in hundreds of data points or fewer, demonstrating its data efficiency. We have also compared Bayesian-SINDy to \texttt{SparseBayes}, which is the foundation of previous Bayesian extensions to SINDy, and demonstrated the importance of considering how noise propagates through the nonlinear candidate functions and the variation of noise level over time. 

Although Bayesian-SINDy has significantly improved data efficiency, its resilience to noise can be further improved. In particular, the cubic oscillator example has revealed the shortcomings of the Gaussian approximation, which limited the noise level Bayesian-SINDy can tolerate. Nevertheless, the Gaussian approximation is found to be broadly valid in other examples, resulting in robust learning performance in real-life data such as the Lynx-Hare population dynamics. To improve the proposed method further, one can seek a more accurate approximation of the noise distribution as it propagates through the library.

Aside from improving the approximation of the noise distribution, much work can be done to expand the current methodology. 
For example, while SINDy can be extended to find PDEs \cite{Rudy2017}, Bayesian-SINDy can also be extended in the same way. 
Another possible extension is the possibility to combine multiple dataset with different parameters. 
While this study has primarily focused on retrieving desired equations from homogeneous datasets in which data are synthesized from the same unknown parameters, our methodology can also combine multiple datasets that share the same underlying dynamics but with different unknown parameters. 
The original SINDy-based method can train parameterised models from datasets of different parameters if the parameters are known or measured as part of the dataset. However, fixing or knowing all parameters in real-world data is typically difficult or impossible to achieve. For example, in modelling cell motility, a population of cells may share a common model for their motility, yet each cell might be parameterised differently and it is difficult to accurately measure each parameter before learning the model. Because the evidence is derived by marginalising the parameters, it is independent of the specific parameter value and provides a mathematical framework to combine multiple datasets with varying unknown parameters.

Another transformative application of Bayesian-SINDy is its easy extension to quantify the information content in each data point, as we have briefly demonstrated in \S\ref{subsec:Active}. By considering the information gain per data point, one can optimise the experimental design to maximise the information gain per data point, thereby significantly improving data efficiency. This is particularly useful in real-life applications such as system identification in control, where the goal is to learn about the system in the most data-efficient manner.

Lastly, it would also be interesting to benchmark Bayesian-SINDy with other recent advancement in SINDy, such as the ensembling technique \cite{Fasel2022} and the denoising method based on automatic differentiation \cite{Kaheman2022}. 
In particular, recent work has established an open standard for benchmarking different sparse system identification methodologies using chaotic systems \cite{Kaptanoglu2023}. 
While we do not expect the Bayesian-SINDy to be the most noise-robust or most accurate in all scenarios, its relatively cheap computational cost and excellent data efficiency make it more suitable for real-time applications such as model identification in control.
Moreover, given Bayesian-SINDy's rigorous justification for model selection under the Bayesian framework, it would be interesting to investigate how this theoretical justification may or may not help its performance in the benchmark compared to other methodologies.

\section*{Acknowledge, contributino and data}
The Bayesian-SINDy method is available at \url{https://github.com/llfung/B-SINDy}. Contributions: L.F.: conceptualisation, formal analysis, methodology, investigation, writing, visualization; U.F.: supervision, validation, review and editing; M.P.J.: conceptualisation, investigation, supervision, review and editing. L.F. acknowledge the support of Research Fellowship from Peterhouse, Cambridge.

\section*{Appendix A. Formula for the library of variance}
To calculate the variances of each column of the library of polynomial terms, we follow the method of Taylor expansion for the moments of functions of random variables \cite{Wolter1985}.

Given a random variable \(X\) with Gaussian distribution \(X \sim \mathcal{N}(\mu,\sigma_x^2)\), the moments of the function \(f(X)\) can be calculated by the Taylor expansion of the function around the mean \(\mu\). In particular, the power function \(f(X) = X^n\) has the following moments:
\begin{equation}
  E[X^n] = \sum_{i=0}^{n} \frac{\sigma_x^i}{i!}\frac{n!}{(n-i)!} \mu^{n-i} a_i,
\end{equation}
where the coefficients \(a_i\) are given by
\begin{equation}
  a_i = \begin{cases}
    \prod_{j=1}^{i/2} (2j-1) & \text{if } i \text{ is even} \\
    0 & \text{if } i \text{ is odd}
  \end{cases}
\end{equation}
and 
\begin{equation}
  \sigma^2_{x^n} = \text{Var}[X^n] = E[X^{2n}] - (E[X^n])^2.
\end{equation}
To calculate the variance of the weighted sum or product of two independent random variables $X_1$ and $X_2$, we use the formula\begin{equation}
  \sigma^2_{a x_1+ b x_2}=\text{Var}[a X_1 + b X_2] = a^2\text{Var}[X_1] + b^2\text{Var}[X_2]
\end{equation}
and
\begin{equation}
  \sigma^2_{x_1x_2}=\text{Var}[X_1X_2] = E[X_1^2]E[X_2^2] - (E[X_1X_2])^2.
\end{equation}

\section*{Appendix B. Selection of hyperparameters values}
Section \ref{sec:examples} compares the learning performances of SINDy (STLS), Bayesian-SINDy and \texttt{SparseBayes} in four examples. To ensure a fair competition between algorithms, we tuned the hyperparameters of each algorithm to maximise their performance in each given example. The resulting hyperparameter values used in each example are summarised in the tables \ref{tbl:hyperparam} and \ref{tbl:lambda_hyperparam}.

\begin{table}[h!]
  \centering
  \caption{Hyperparameters used in each example in \S\ref{sec:examples}.}
  \label{tbl:hyperparam}
  \begin{tabular}{c|ccc}
                              & Prior Variance          & Weak                       & Finite     \\
                 & $\alpha^{-1}$ (B-SINDy) & Formulation                & Difference \\ \hline
Van der Pol      & $10^2$                  & $\phi=(t^2-1)^2$, 9 points & 8th order  \\
Cubic Oscillator & $1^2$                     & $\phi=(t^2-1)^4$, 6 points &            \\
Lorenz           & $25^2$                  &                            & 12th order \\
Lynx-Hare        & $10^2$                  &                            & 8th order 
  \end{tabular}%
\end{table}

For SINDy (STLS), the threshold parameter $\lambda$ is tuned in increments of the first significant figure to maximise the overall success rate at each noise level in the synthetic examples and tuned to recreate the Lotka-Volterra equation with the least number of spurious terms in the Lynx-Hare example (table \ref{tbl:lambda_hyperparam}). For Bayesian-SINDy, the choice for the prior variance $\alpha^{-1}$ is chosen to be of similar order to the expected parameter value. The values of the prior variance represent prior knowledge of the order of magnitude of the parameter value. In practice, this rough knowledge of the parameters' order of magnitude can usually be obtained from simple scaling arguments or physical intuition. Meanwhile, the \texttt{SparseBayes} algorithm has no hyperparameter that requires tuning, as both $\boldsymbol\alpha$ and $\boldsymbol\beta$ are optimised by the algorithm.

\begin{table}[h]
  \centering
  \caption{Optimal values for thresholding parameter $\lambda$ in SINDy (STLS) in each example in \S\ref{sec:examples}.}
  \label{tbl:lambda_hyperparam}
\begin{tabular}{c|cc}
                 & Noise level $\sigma_x$ & Threshold Parameter $\lambda$ \\ \hline
Van der Pol      & $0.1$                  & $0.4$                         \\
                 & $0.2$                  & $0.4$                         \\ \hline
Cubic Oscillator & $0.005$                & $0.04$                        \\
                 & $0.01$                 & $0.06$                        \\ \hline
Lorenz           & $0.025$                & $0.2$                         \\
                 & $0.05$                 & $0.3$                         \\
                 & $0.1$                  & $0.4$                         \\
                 & $0.2$                  & $0.5$                         \\ \hline
Lynx-Hare        & N/A                    & $0.025$                      
\end{tabular}
\end{table}

The choices between weak formulation and finite difference are made according to the sampling frequency of the data compared to the typical frequency of the data. As per the discussion in \S\ref{sec:WeakVSFD}, the weak formulation is preferred if the sampling frequency is found to be suitably higher than the typical frequency of the dynamics. Note that in practice, the application of weak formulation or finite difference should be decided according to the signal in the data before the learning takes place. Here, the weak formulation is used in the Cubic Oscillator example because it does not significantly distort the signal. In other examples, the finite difference is chosen because crucial high-frequency signals are filtered out when the weak formulation is used.

\vskip2pc

\bibliographystyle{RS} 
\bibliography{Library} 


\end{document}